\documentclass[11pt]{article}
\usepackage[table]{xcolor}
\definecolor{light-gray}{gray}{0.90}
\usepackage{amsmath}
\usepackage{amsthm}
\usepackage{enumerate}
\usepackage{graphicx,psfrag,epsf}
\usepackage{amssymb}
\usepackage{hyperref}
\usepackage{diagbox}
\usepackage{geometry}
\usepackage[utf8]{inputenc}
\usepackage{float}
\usepackage{todonotes}
\usepackage[title]{appendix}
\usepackage{algorithm}
\usepackage{algpseudocode}
\usepackage[none]{hyphenat}
\usepackage{titlesec}
\usepackage[official]{eurosym}
\usepackage{mathtools}
\usepackage{natbib}
\usepackage{caption}
\usepackage{subcaption}
\usepackage{color}
\usepackage{algorithm}
\usepackage{algorithmicx}
\usepackage{algpseudocode}
\usepackage{color}
\usepackage{extarrows}
\usepackage{booktabs}
\usepackage{makecell}
\usepackage{graphicx}
\usepackage{bm}

\renewcommand{\vec}[1]{\bm #1}
\newcommand{\data}{{X}^k}
\newcommand{\dataValue}{{x}^k}
\newcommand{\repvec}[1]{{\mathbb \langle} #1 {\mathbb \rangle}}

\setuptodonotes{fancyline, color=blue!30}

\numberwithin{equation}{section}

\newcolumntype{C}[1]{>{\centering\arraybackslash}m{#1}}
\newcolumntype{R}[1]{>{\raggedleft\arraybackslash}m{#1}}
\definecolor{mygray}{gray}{0.9}

\newtheorem{proposition}{Proposition}

\newtheorem{theorem}{Theorem}
\newtheorem{corollary}{Corollary}

\theoremstyle{plain}
\newtheorem{definition}{Definition}
\newtheorem{lemma}{Lemma}

\newcommand{\Scond}{\ensuremath{{S}_{\textsc{cond}}}}
\newcommand{\Sgro}[1]{\ensuremath{{S}_{\textsc{gro}(#1)}}}
\newcommand{\Sripr}{\ensuremath{{S}_{\textsc{gro}(\cM)}}}
\newcommand{\Smix}{\ensuremath{{S}_{\textsc{gro}(\textsc{iid})}}}
\newcommand{\Spseudo}{\ensuremath{{S}_{\textsc{pseudo}(\cM)}}}
\newcommand{\Sappr}{\ensuremath{{S}_{\textsc{appr}}}}

\newcommand{\cH}{\ensuremath{{\cal H}}}
\newcommand{\cP}{\ensuremath{{\cal P}}}

\newcommand{\cX}{\ensuremath{{\cal X}}}
\newcommand{\cM}{\ensuremath{{\cal M}}}
\newcommand{\E}{\textit{e}}

\newtheorem{examplehidden}{Example}
\newenvironment{example}{
\begin{examplehidden}
\em
}
{\end{examplehidden}}

\title{\texttt{E}-values for $k$-Sample Tests With Exponential Families
    }
\author{Yunda Hao \\ CWI\thanks{Centrum Wiskunde \& Informatica (CWI) is the national research institute for mathematics and computer science in the Netherlands, located in Amsterdam.} \\ \and Peter Gr\"unwald\thanks{Corresponding author: pdg@cwi.nl} \\ CWI and Leiden University \\  \and Tyron Lardy \\ Leiden University \\  \and  Long Long  \\ Yiyang \\
 \and Reuben Adams \\ University College London }
\begin{document}
\bibliographystyle{plainnat}
\maketitle

\begin{abstract}
We develop and compare \E-variables for testing whether $k$ samples of data are drawn from the same distribution, the alternative being that they come from different elements of an  exponential family.
We consider the GRO (growth-rate optimal) \E-variables for (1) a `small' null inside the same exponential family, and  (2) a `large' nonparametric null, as well as (3) an  \E-variable arrived at by conditioning on the sum of the sufficient statistics.
(2) and (3) are efficiently computable, and extend ideas from \cite{turner2021safe} and \cite{Wald1947} respectively from Bernoulli to general exponential families. 
We provide theoretical and simulation-based comparisons of these \E-variables in terms of their logarithmic growth rate, 
and find that for small effects all four \E-variables behave surprisingly similarly; for the Gaussian location and Poisson families, e-variables (1) and (3) coincide; for Bernoulli, (1) and (2) coincide; but in general, 
whether (2) or (3) grows faster under the alternative is family-dependent. 
We furthermore discuss algorithms for numerically approximating (1).
\end{abstract}




\section{Introduction}\label{sec:introduction}
E-variables (and the value they take, the {\em \E-value\/}) provide an alternative to p-values that is inherently more suitable for testing under optional stopping and continuation, and that lies at the basis of {\em anytime-valid\/} confidence intervals that can be monitored continuously \citep{grunwald2019safe,VovkW21,Shafer21,ramdas2022game,henzi2021valid,Grunwald23}.  
While they have their roots in  the work  on anytime-valid testing by H. Robbins and students (e.g. \citep{darling1967confidence}), they have begun to be investigated in detail for composite null hypotheses only very recently. 
E-variables can be associated with a natural notion of optimality, called GRO (growth-rate optimality), introduced and studied in detail by \cite{grunwald2019safe}. 
GRO may be viewed as an analogue of the uniformly most powerful test in an optional stopping context.
In this paper, we develop GRO and near-GRO \E-variables for a classical statistical problem: parametric $k$-sample tests.
Pioneering work in this direction appears already in \cite{Wald1947}: as we explain in Example~\ref{ex:wald}, his SPRT for a sequential test of two proportions can be re-interpreted in terms of \E-values for Bernoulli streams. Wald's \E-values are not optimal in the GRO sense --- GRO versions were derived only very recently by  \cite{turner2021safe,TurnerG22}, but again only for Bernoulli streams. 
Here we develop \E-variables for the case that the alternative is 
associated with an arbitrary but fixed exponential family, $\cM$, with data in each of the $k$ groups sequentially sampled from a different distribution in that family. 
We mostly consider tests against the null hypothesis, denoted by $\cH_0(\cM)$ that states that outcomes in all groups are i.i.d. by a single member of $\cM$.
We develop the GRO \E-variable $\Sripr$ for this null hypothesis, but it is not efficiently computable in general.
Therefore, we introduce two more tractable \E-variables: $\Smix$ and $\Scond$.
The former is defined as the GRO \E-variable, for the much larger null hypothesis that the $k$ groups are i.i.d. from an arbitrary distribution, denoted by $\cH_0(\textsc{iid})$: since an e-variable relative to a null hypothesis $\cH_0$ is automatically an e-variable relative to any null that is a subset of $\cH_0$, $\Smix$ is automatically also an e-variable relative to $\cH_0(\cM)$. Whenever below we refer to `the null', we mean the smaller $\cH_0(\cM)$. The use of $\Smix$ rather than $\Sripr$ for this null, for which it is not GRO, is justifiable by ease of computation and robustness against misspecification of the model $\cM$.
However, exactly this robustness might also cause it to be too conservative when $\cM$ is well-specified.
The third  \E-variable we consider, $\Scond$, does not have any GRO status, but is specifically tailored to $\cH_0(\cM)$, so that it might still be better than $\Smix$ in practice.
Finally, we introduce a pseudo-\E-variable $\Spseudo$, which coincides with $\Sripr$ whenever the latter is easy to compute; in other cases it is not a real \E-variable, but it is still highly useful for our theoretical analysis. 

\paragraph{Results}
Besides defining $\Sripr$, $\Smix$ and $\Scond$ and proving that they achieve what they purport to, we analyze their behaviour both theoretically and by simulations. Our main theoretical results, Theorem~\ref{Taylor-approximation} and ~\ref{Taylor-approximation Scond} reveal some surprising facts: for any exponential family, the four types of (pseudo-) \E-variables achieve almost the same growth rate under the alternative, hence are almost equally good, whenever the `distance' between null and alternative is sufficiently small. 
That is, suppose that the (shortest) $\ell_2$-distance between the $k$ dimensional parameter of the alternative and the parameter space of the null is given by $\delta$. 
Then for any two of the aforementioned \E-variables $S,S'$, we have $\mathbb{E}[\log S-\log S']=O(\delta^4)$, where the expectation is taken under the alternative. 
Here, $\mathbb{E}[\log S]$ can be interpreted as the growth rate of $S$, as explained in Section~\ref{sec:setting}.

While $\Smix$ and $\Scond$ are efficiently computable for the families we consider, this is generally not the case for $\Sripr$, since to compute it we need to have access to the {\em reverse information projection\/} (RIPr; \citep{li1999estimation,grunwald2019safe}) of a fixed simple alternative to the set $\cH_0(\cM)$. 
In general, this is a convex combination of elements of $\cH_0(\cM)$, which can only be found by numerical means. 
Interestingly, we find that for three families, Gaussian with fixed variance, Bernoulli and Poisson, 
the RIPr is attained at a single point (i.e. a mixture putting all its mass on that point) that can be efficiently computed.
Furthermore, in these cases $\Sripr$ coincides with one of the other \E-variables ($\Smix$ for Bernoulli, $\Scond$ for Gaussian and Poisson). For other exponential families, for $k=2$, we approximate the RIPr and hence $\Sripr$ using both an algorithm proposed by Li (\citeyear{li1999estimation}) and a brute-force approach. 
We find that we can already get an extremely good approximation of the RIPr with a mixture of just {\em two\/} components. 
This leads us to conjecture that perhaps the deviation from the RIPr is just due to numerical imprecision and that the actual RIPr really can be expressed with just two components. 
The theoretical interest of such a development notwithstanding, we advise to use
$\Scond$ or $\Smix$ rather than $\Sripr$ for practical purposes whenever more than one component is needed for the RIPr, as their growth rates are not much worse, and they are much easier to compute.
If furthermore robustness against misspecification of the null is required, then $\Smix$ is the most sensible choice.

\paragraph{Method: Restriction to Single Blocks and Simple Alternatives} The main interest of \E-variables is in analyzing sequential, anytime-valid settings: the data arrives in $k$ streams corresponding to $k$ groups, and we may want to stop or continue sampling at will (optional stopping); for example, we only stop when the data looks sufficiently good; or we stop unexpectedly, because we run out of money to collect new data. 
Nevertheless, in this paper we focus on what happens in a single {\em block}, i.e. a vector $\data=(X_1, \ldots, X_k)$, where each $X_j$ denotes a single outcome in the $j$-th stream. 
By now, there are a variety of papers (see e.g. \cite{grunwald2019safe,ramdas2022game,turner2021safe}) that explain how \E-variables defined for such a single block can be combined by multiplication to yield e-processes (in our context, coinciding with {\em nonnegative supermartingales\/}) that can be used for testing the null with optional stopping if blocks arrive sequentially --- that is, one observes one outcome of each sample at a time. Briefly, one multiplies the \E-variables and at any time one intends to stop, one rejects the null if the product of \E-values observed so-far exceeds $1/\alpha$ for pre-specified significance level $\alpha$. This gives an {\em anytime-valid\/} test at level $\alpha$: irrespective of the stopping rule employed, the Type-I error is guaranteed to be below $\alpha$. Similarly, one can extend the method to design {\em anytime-valid confidence intervals}  by inverting such tests, as described in detail by \cite{ramdas2022game}. This is done for the $2$-sample test with Bernoulli data by \cite{TurnerG22}; their inversion methods are extendable to the general exponential family case we discuss here.  Thus, we refer to the aforementioned papers for further details and restrict ourselves here to the 1-block case.  
Also, \cite{turner2021safe,TurnerG23} describe how one can adapt an e-process for data arriving in blocks to general streams in which the $k$ streams do not produce data points at the same rate; we briefly extend their explanation to the present setting in Appendix~\ref{sec:streaming}. 
Finally, we mainly restrict to the case of a simple alternative, i.e. a single member of the exponential family under consideration. 
While this may seem like a huge restriction, extension from simple to composite alternatives (e.g. the full family under consideration) is straightforward using the {\em method of mixtures\/} (i.e. Bayesian learning of the alternative over time) and/or the plug-in method. We again refer to \cite{grunwald2019safe,ramdas2022game} for detailed explanations, and \cite{turner2021safe} for an explanation in the 2-sample Bernoulli case, and restrict here to the simple alternative case: all the `real' difficulty lies in dealing with composite null hypotheses, and that, we do explicitly and exhaustively in this paper. 
    
\paragraph{Related Work and Practical Relevance} As indicated, this paper is a direct (but far-reaching) extension of the papers \cite{turner2021safe,TurnerG22} on 2-sample testing for Bernoulli streams as well as Wald's (\citeyear{Wald1947}) sequential two-sample test for proportions to streams coming from an exponential family. 
There are also {\em nonparametric\/} sequential \citep{lheritier2018sequential} and anytime-valid 2-sample tests \citep{balsubramani2015sequential,Pandeva22} that tackle a somewhat different problem. 
They work under much weaker assumptions on the alternative (in some versions the samples could be arbitrary high-dimensional objects such as pictures and the like). 
The price to pay is that they will need a much larger sample size before a difference can be detected. 
Indeed, while our main interest is theoretical (how do different \E-variables compare? in what sense are they optimal?), in settings where data are expensive, such as randomized clinical trials, the methods we describe here can be practically very useful: they are exact (existing methods are often based on chi-squared tests, which do not give exact Type-I error guarantees at small sample size), they allow for optional stopping, and they need small amounts of data due to the strong parametric assumptions for the alternative. 
As a simple illustration of the practical importance of these properties,
we refer to the recent SWEPIS study \citep{wennerholm2019induction} which was stopped early for harm. 
As demonstrated by \cite{turner2021safe}, if an anytime-valid two-sample test had been used in that study, substantially stronger conclusions could have been drawn. 

We also mention that  $k$-sample tests can be viewed as independence tests (is the outcome independent of the group it belongs to?) and as such this paper is also related to recent papers on \E-values and anytime-valid tests for conditional independence testing \citep{HenziLardy22,Shaer2022,Duan2022}. Yet,  the setting studied in those papers is quite different in that they assume the covariates (i.e. indicator of which of the $k$ groups the data belongs to) to be i.i.d.
\paragraph{Contents}
In the remainder of this introduction, we fix the general framework and notation and we briefly recall how \E-variables are used in an anytime-valid/optional stopping setting.  
In Section~\ref{sec:four} we describe our four (pseudo-) \E-variables in detail, and we provide preliminary results that characterize their behaviour in terms of growth rate. 
In Section~\ref{Theoretical_results} we provide our main theoretical results which show that, for all regular exponential families, the expected growth of the four types of \E-variables is of surprisingly small order $\delta^4$ if the parameters of the alternative are at $\ell_2$-distance $\delta$ to the parameter space of the null. 
In Section~\ref{sec:specific} we give more detailed comparisons for a large number of standard exponential families (Gaussian, Bernoulli, Poisson, exponential, geometric, beta), 
including simulations that show what happens if $\delta$ gets larger.
Section~\ref{sec:simulations} provides some additional simulations about the RIPr. All proofs, and some additional simulations, are in the appendix. 

\subsection{Formal Setting}\label{sec:setting}
Consider a regular one-dimensional exponential family $\cM = \{P_{\mu}: \mu \in \mathtt{M}\}$ given in its mean-value parameterization (see e.g. \citep{BarndorffNielsen78} for more on  definitions and for all the proofs of all standard results about exponential families that are to follow). Each member of the family is a distribution 
for some random variable $U$, taking values in some set ${\cal U}$, with density $p_{\mu;[U]}$ relative to some underlying measure $\rho_{[U]}$ which, without loss of generality, can be taken to be a probability measure. For regular exponential families, $\mathtt{M}$ is an open interval in ${\mathbb R}$ and $p_{\mu;[U]}$ can be written as:
\begin{equation}\label{ExponentialFormulaPre}
p_{\mu;[U]}(U) = \exp\left(\lambda(\mu) \cdot t(U) - A(\lambda(\mu)) \right),
\end{equation}
where $\lambda(\mu)$ maps mean-value $\mu$ to canonical parameter $\beta$. We then have $\mu = \mathbb{E}_{P_{\mu}}[t(U)]$, where  $t(U)$ is a measurable function of $U$  and $A(\beta)$ is the log-normalizing factor. The measure $\rho_{[U]}$ induces a corresponding (marginal) measure $\rho := \rho_{[X]}$ on the {\em sufficient statistic\/}
$X:= t(U)$, and similarly the  density (\ref{ExponentialFormulaPre}) induces  a corresponding density $p_{\mu} := p_{\mu;[X]}$ on $X$, i.e. we have 
\begin{equation}\label{ExponentialFormula}
p_{\mu}(X) := p_{\mu;[X]}(X) = \exp\left(\lambda(\mu) \cdot X - A(\lambda(\mu)) \right).
\end{equation}
All \E-variables that we will define can be written in terms of   the induced measure and density of the sufficient statistic of $X$; in other words, we can without loss of generality act as if our family is {\em natural}. Therefore, from now on we simply assume that we observe data in terms of their sufficient statistics $X$ rather than the potentially more fine-grained $U$, and will be silent about $U$; 
for simplicity we thus abbreviate $p_{\mu; [X]}$ to $p_{\mu}$ and $\rho_{[X]}$ to $\rho$. Note that exponential families are more usually defined with a carrier function $h(X)$ and $\rho$ set to Lebesgue or counting measure; we cover this case by absorbing $h$ into $\rho$, which we do not require to be Lebesgue or counting. 

The data comes in as a block $X^k = (X_1, \ldots, X_k) \in \cX^k$, where $\cX$ is the support of $\rho$. To calculate our \E-values we only need to know $\data\in \cX^k$, and under the alternative hypothesis, all $X_j$, $j=1\dots k$ are distributed according to some element $P_{\mu_j}$ of $\cM$. 
In our main results we take the alternative hypothesis to be {\em simple}, i.e. we assume that $\vec{\mu} = (\mu_1, \ldots, \mu_k)\in \mathtt{M}^k$ is fixed in advance. 
The alternative hypothesis is thus given by
$$\text{simple } \mathcal{H}_1: X_1 \sim P_{\mu_1}, X_2 \sim P_{\mu_2}, \dots, X_k \sim P_{\mu_k} \  \text{independent}.$$
Note that we will keep $\vec{\mu}$ fixed throughout the rest of this section and Section \ref{sec:four}.
This is without loss of generality as $\vec{\mu}$ is defined as an arbitrary element of $\mathtt{M}^k$, so that all results stated for $\vec{\mu}$ hold for any element of $\mathtt{M}^k$. 
The extension to composite alternatives by means of the method of mixtures or the plug-in method is straightforward, and done in a manner that has become standard for \E-value based testing \citep{ramdas2022game}. 

Our null hypothesis is directly taken to be composite, for as regards the null, the composite case is inherently very different from the simple case \citep{ramdas2022game,grunwald2019safe}. 
It expresses that the $\data$ are identically distributed.
We shall consider various variants of this null hypothesis, all composite: 
let ${\cal P}$ be a set of distributions on ${\cal X}$, then the null hypothesis {\em relative to ${\cal P}$}, denoted $\cH_0(\cP)$, is defined as
$$\text{composite } \mathcal{H}_0(\cP) : 
X_1 \sim P, X_2 \sim P, \ldots, X_k \sim P 
\hspace{.25cm} \text{i.i.d. for some } P  \in \cP.
$$
Our most important instantiation for the null hypothesis will be $\cH_0= \cH_0(\cM)$ for the same exponential family $\cM$ from which the alternative was taken; then $\cH_0(\cM)$ is a one-dimensional parametric family expressing that the $X_i$ are i.i.d. from $P_{\mu_0}$ for $\mu_0 \in \mathtt{M}$. 
Still, we will also consider $\cH_0= \cH_0(\cP)$ where $\cP$ is the much larger set of {\em all\/} distributions on $\cX$. 
Then the null simply expresses that the $\data$ are i.i.d.; we shall abbreviate this null to $\cH_0(\textsc{iid})$. 
Finally we sometimes consider $\cH_0 = \cH_0(\cM')$ where $\cM'\subset \cM$ is a subset of $P_{\mu} \in \cM$ with $\mu \in \mathtt{M}'$ for some sub-interval $\mathtt{M'} \subset \mathtt{M}$.
The statistics that we use to gain evidence against these null hypotheses are \E-variables. 
\begin{definition}
We call any nonnegative random variable $S$ on a sample space $\Omega$ (which in this paper will always be $\Omega = \cX^k$) an {\em \E-variable relative to $\cH_0$\/} if it satisfies  
\begin{equation}\label{eq:evar}
\text{ for all } P \in \cH_0: \qquad \mathbb{E}_P[S] \leq 1.
\end{equation}
\end{definition}
\subsection{{The GRO E-variable for  General \texorpdfstring{${\cal H}_0$}{H\_0}}}\label{sec:gro}
In general, there exist many \E-variables for testing any of the null hypotheses introduced above. 
Each \E-variable $S$ can in turn be associated with a growth rate, defined by $\mathbb{E}_{P_{\vec{\mu}}}[\log S]$.
Roughly, this can be interpreted as the (asymptotic) exponential growth rate one would achieve by using $S$ in consecutive independent experiments and multiplying the outcomes if the (simple) alternative was true (see e.g.~\citep[Section 2.1]{grunwald2019safe} or~\citep{kelly1956ANI}). 
The Growth Rate Optimal (GRO) \E-variable is then the \E-variable with the greatest growth rate among all \E-variables. The central result (Theorem 1) of \cite{grunwald2019safe} states that, under very weak conditions, GRO \E-variables take the form of likelihood ratios between the alternative and the {\em reverse information projection\/} \citep {li1999estimation} of the alternative onto the null. We instantiate their Theorem 1 to our setting by providing Lemma~\ref{lem:safe} and~\ref{lem:gro}, both special cases of their Theorem 1. 
Before stating these, we need to introduce some more notation and definitions.
For $\vec{\mu} =(\mu_1, \ldots, \mu_k)$ we use the following notation: 
$$p_{\vec{\mu}}(\data) := \prod\limits_{i=1}^k p_{\mu_i}(X_i).$$
Whenever in this text we refer to KL divergence $D(Q \| R)$, we refer to measures $Q$ and $R$ on $\cX^k$. Here $Q$ is required to be a probability measure, while $R$ is allowed to be a sub-probability measure, as in \citep{grunwald2019safe}. 
A {\em sub-\/} probability measure $R$ on $\cX^k$ is a measure that integrates to 1 or less, i.e $\int_{x \in \cX} d R(x) \leq 1$. 

The following lemma follows as a very special case of 
Theorem 1 (simplest version) of  \cite{grunwald2019safe}, when instantiated to our $k$-sample testing set-up: 
\begin{lemma}\label{lem:safe}
Let $\cP$ be a set of probability distributions on $\cX^k$ and let $\textsc{conv}(\cP)$ be its convex hull.
Then there exists a sub-probability measure $P^*_0$ with density $p^*_0$ such that 
\begin{equation}\label{eq:kl_min}
D(P_{\vec{\mu}} \| P^*_0) = \inf_{P \in \textsc{conv}(\cP)} D(P_{\vec{\mu}} \| P).
\end{equation}  
$P^*_0$ is called the {\em reverse information projection (RIPr)} of $P_{\vec{\mu}}$ onto $\textsc{conv}(\cP)$.
\end{lemma}
Clearly, if $P^*_0 \in \textsc{conv}(\cP)$ (the minimum is achieved) then $P^*_0$ is a probability measure, i.e. integrates to exactly one.
We show that this happens for certain specific exponential families in Section~\ref{sec:specific}.
However, in general we can neither expect the minimum to be achieved, nor the RIPr to integrate to one.
%
Lemma~\ref{lem:gro} below, again a special case of \cite[Theorem 1]{grunwald2019safe}, shows that the RIPr characterizes the GRO \E-variable, and explains the use of the term \textsc{gro} in the definition below. 
\begin{definition}
 $\Sgro{\cP}$ is defined as
\begin{equation}\label{eq:Sgro}
\Sgro{\cP} := \frac{p_{\vec{\mu}}(\data)}{p^{*}_0(\data)}
\end{equation}
where $p^*_0$ is the density of the
RIPr of $P_{\vec{\mu}}$ onto $\textsc{conv}(\cP)$.
\end{definition}
\begin{lemma}\label{lem:gro}
For every set of distributions $\cP$ on $\cX$, $\Sgro{\cP}$ is an \E-variable for $\cH_0(\cP)$. Moreover, it is the
GRO {\em (Growth-Rate-Optimal)\/} \E-variable for $\cH_0(\cP)$, i.e. it essentially uniquely achieves
$$
\sup_{S} {\mathbb{E}}_{P_{\vec{\mu}}} [\log S]
$$
where the supremum ranges over all \E-variables for $\cH_0(\cP)$. 
\end{lemma}
Here, essential uniqueness means that any other GRO \E-variable must be equal to $\Sgro{\cP}$ with probability $1$ under $P_{\vec{\mu}}$.
This in turn implies that the measure $P^*_0$ is in fact unique, as members of regular exponential families must have full support. 
Thus, once we have fixed our alternative and defined our null as $\cH_0(\cP)$ for some set of distributions $\cP$ on $\cX$, the optimal (in the GRO sense) \E-variable to use is the $\Sgro{\cP}$ \E-variable as defined above.

\section{The Four Types of E-variables}
\label{sec:four}
In this section, we define our four types of \E-variables; the definitions can be instantiated to any underlying 1-parameter exponential family. More precisely, we define three `real' \E-variables $\Sripr, \Scond, \Smix$ and one `pseudo-\E-variable' $\Spseudo$, a variation of $\Sripr$ which for some exponential families is an \E-variable, and for others is not.  
\subsection{The GRO E-variable for \texorpdfstring{$\cH_0(\cM)$}{H\_0(M)} and the pseudo \E-variable}
We now consider the GRO \E-variable for our main null of interest, $\cH_0(\cM)$. 
In practice, for some exponential families $\cM$, the infimum over $\textsc{conv}(\cM)$ in~\eqref{eq:kl_min} is actually achieved for some $P_{\mu^*_0} \in \cM$. In this {\em easy\/} case we can determine $\Sripr$  analytically (this happens if $\Sripr= \Spseudo$, see below). For all other $\cM$, i.e. whenever the infimum is not achieved at all or is in $\textsc{conv}(\cM) \setminus \cM$,  we do not know if $\Sripr$ can be determined analytically. In this {\em hard\/} case  will numerically approximate it by $\Sripr'$ as defined below.
First, for a fixed parameter $\mu_0 \in \texttt{M}$ we define the vector $\repvec{\mu_0}$ as the vector indicating the distribution on $\cX^k$ with all parameters equal to $\mu_0$:
\begin{equation}
    \label{eq:repvec}
    \repvec{\mu_0} := (\mu_0, \ldots, \mu_0) \in \texttt{M}^k
\end{equation}
Next, with 
$W$ a distribution on $\texttt{M}$, we define 
\begin{equation}\label{eq:bayesmarginal}
p_{W} := \int p_{\repvec{\mu_0}}(\data) dW (\mu_0)
\end{equation}
to be the Bayesian marginal density obtained by marginalizing over distributions in $\cH_0(\cM)$ according to $W$. 
Clearly, if $W$ has finite support then the corresponding distribution $P_W$ has $P_W \in \textsc{conv}(\cM)$. We now set 
$$
\Sripr' := \frac{p_{\vec{\mu}}(\data)}{p_{W'_0}(\data)}
$$
where $W'_0$ is chosen so that  $p_{W'_0}$ is within a small  $\epsilon$ of achieving the minimum in~\eqref{eq:kl_min}, 
i.e. $D(P_{\mu_1,\dots,\mu_k}\|P'_{W_0}) = \inf_{P\in \textsc{conv}(\cM)} D(P_{\mu_1,\dots,\mu_k}\|P)+\epsilon'$ for some $0 \leq \epsilon' < \epsilon$.
Then, by Corollary 2 of~\citet{grunwald2019safe}, $\Sripr'$ will {\em not\/} be an \E-variable unless $\epsilon' = 0$, 
but in each case (i.e. for each choice of $\cM$) we verify numerically that $\sup_{\mu_0 \in \texttt{M}} \mathbb{E}_{P_{\mu_0,\ldots, \mu_0}}[S] = 1 + \delta$ for negligibly small $\delta$, i.e. $\delta$ goes to $0$ quickly as $\epsilon'$ goes to $0$. 
We return to the details of the calculations in Section~\ref{sec:simulations}. 

We now consider the `easy' case in which $P^*_0 = P_{\repvec{\mu^*_0}}$ for some  $\mu^*_0 \in \mathtt{M}$. Clearly, we must have  $\mu_0^{*} := \arg\min_{\mu_0 \in \mathtt{M}} {D}(P_{\vec{\mu}} \| P_{\repvec{\mu_0}})$. An easy calculation shows that then 
\begin{equation}\label{eq:mu0star}
\mu_0^{*} = \frac{1}{k} \sum\limits_{i=1}^k \mu_i.
\end{equation}
\begin{definition}\label{DefSpseudo}
$\Spseudo$ is defined as
$$\Spseudo := \frac{p_{\vec{\mu}}(\data)}{p_{\repvec{\mu_0^*}}(\data)}.$$
\end{definition}
$\Spseudo$ is not always a real \E-variable relative to $\cH_0(\cM)$, which explains the name `pseudo'.
Still, it will be very useful as an auxiliary tool in Theorem~\ref{Taylor-approximation} and derivations. Note that, if it is an \E-variable then we know that it is equal to $\Sripr$: 
\begin{proposition}\label{Speudo=Sripr}
$\Spseudo$ is an \E-variable for $\cM$ iff $\Spseudo = \Sripr$.
\end{proposition}
The proposition above does not give any easily verifiable condition to check whether $\Spseudo$ is an \E-variable or not. The following proposition does provide a condition which is sometimes easy to check (and which we will heavily employ below).
With $\mu^*_0$ as in (\ref{eq:mu0star}), define
$$
f(\mu_0) := \sum\limits_{i=1}^k \text{\sc var}_{P_{\mu_i+\mu_0-\mu_0^{*}}}[X] - k\text{\sc var}_{P_{\mu_0}}[X].
$$
\begin{proposition}\label{SpseudoAnE-value}
If $f(\mu^*_0)> 0$, then $\Spseudo$ is not an \E-variable. If $f(\mu^*_0) < 0$, then there exists an interval  $\mathtt{M'} \subset \mathtt{M}$  with $\mu_0^*$ in the interior of $\mathtt{M'}$ so that $\Spseudo$ is an \E-variable 
for $\cH_0(\cM')$, where $\cM'= \{P_{\mu}: \mu \in \mathtt{M'}\}$. 
\end{proposition}

\subsection{The GRO E-variable for \texorpdfstring{{\rm  $\cH_0(\textsc{iid})$}}{H\_0(iid)}}
Recall that we defined $\cH_0(\textsc{iid})$ as the set of distributions under which $X_j$, $j=1,\dots k$, are i.i.d. from some arbitrary distribution on $\cX$.
By the defining property of \E-variables, i.e. expected value bounded by one under the null~\eqref{eq:evar}, it should be clear that any \E-variable for $\cH_0(\textsc{iid})$ is also an \E-variable for $\cH_0(\cM)$, since $\cH_0(\cM)\subset \cH_0(\textsc{iid})$. 
In particular, we can also use the GRO \E-variable for $\cH_0(\textsc{iid})$ in our setting with exponential families.  It turns out that this \E-variable, which we will denote as $\Smix$, has a simple form that is generically easy to compute. We now show this: 
\begin{theorem}\label{thm:smix} 
The minimum KL divergence  $\inf_{P\in \textsc{conv}(\cH_0(\textsc{iid}))} D(P_{\vec{\mu}}\| P)$ as in Lemma~\ref{lem:safe}
is achieved by the distribution $P^*_0$ on $\mathcal{X}^k$ with density
\[p^*_0({x}^k)= \prod_{j=1}^k \frac{1}{k} \sum\limits_{i=1}^k p_{\mu_i}(x_j).\] 
Hence, 
$\Smix$, as defined below, is the GRO \E-variable for $\cH_0(\textsc{iid})$. 
\end{theorem}
\begin{definition}\label{Definition_Smix}
$\Smix$ is defined as
\[\Smix := \frac{p_{\vec{\mu}}(\data)}{\prod\limits_{j=1}^k \left(\frac{1}{k} \sum\limits_{i=1}^k p_{\mu_i}(X_j)\right)}.\]
\end{definition}
The proof of Theorem~\ref{thm:smix} extends an argument of~\cite{turner2021safe} for the 2-sample Bernoulli case to the general $k$-sample case. The argument used in the proof does not actually require the alternative to equal the product distribution of $k$ independent elements of an exponential family --- it could be given by the product of $k$ arbitrary distributions. 
However, we state the result only for the former case, as that is the setting we are interested in here. 

\subsection{The Conditional E-variable \texorpdfstring{$\Scond$}{Scond}}
So far, we have defined \E-variables as likelihood ratios between $P_{\vec{\mu}}$ and cleverly chosen elements of either $\cH_0(\cM)$ or $\cH_0(\textsc{iid})$.
We now do things differently by not considering the full original data $X_1,\dots X_k$, but instead conditioning on the sum of the sufficient statistics, i.e. $Z=\sum_{i=1}^k X_i$.
It turns out that doing so actually collapses $\cH_0(\cM)$ to a single distribution, so that the null becomes simple.
That is, the distribution of $\data \mid Z$ is the same under all elements of $\cH_0(\cM)$, as we will prove in Proposition~\ref{prop:scond}. 
This means that instead of using a likelihood ratio of the original data, we can use a likelihood ratio conditional on $Z$, which `automatically' gives an \E-variable.
\begin{definition}
Setting $Z$ to be the random variable $Z:= \sum_{i=1}^k X_i$, $\Scond$ is defined as
$$\Scond := \frac{p_{\vec{\mu}}\left({X}^{k-1} \mid Z \right)}{p_{\repvec{\mu_0}}\left({X}^{k-1} \mid Z \right)},$$
with $\mu_0 \in {\mathtt M}$ and $(X)$ the sufficient statistic as in \eqref{ExponentialFormula}.
\end{definition}

\begin{proposition}\label{prop:scond}
For all $\vec{\mu}' = (\mu'_1,\dots,\mu'_k) \in {\mathtt M}^k$ , we have that $p_{\vec{\mu}'}({x}^{k-1}\mid Z=z)$ depends on $\vec{\mu}'$ only through $\lambda_j := \lambda(\mu_j')-\lambda(\mu_k')$, $j=1,\dots k-1$, i.e. it can be written as a function of $(\lambda_1, \ldots, \lambda_{k-1})$.
As a special case, for all $\mu_0,\mu'_0\in {\mathtt M},$ it holds that $ p_{\repvec{\mu_0}}(\dataValue\mid Z)=p_{\repvec{\mu'_0}}(\dataValue\mid Z)$.
As a direct consequence, $\Scond$ is an \E-variable for $\cH_0(\cM)$, 
\end{proposition}
\begin{example}{\bf [The Bernoulli Model]}\label{ex:wald}
If $\cM$ is the Bernoulli model and $k=2$, then the conditional \E-variable reduces to a ratio between the conditional probability of $(X_1,X_2) \in \{0,1\}^2$ given their sum $Z \in \{0,1,2\}$. 
Clearly, for all $\mu'_1, \mu'_2 \in {\mathtt{M}}= (0,1)$, we have $p_{\mu'_1,\mu'_2}((0,0) \mid Z=0) = p_{\mu'_1,\mu'_2}((1,1) \mid Z=2) = 1$, so $\Scond= 1$ whenever $Z=0$ or $Z=2$, irrespective of the alternative: data with the same outcome in both groups is effectively ignored. 
A non-sequential version of $\Scond$ for the Bernoulli model was analyzed earlier in great detail  by \cite{adams2020safe}.

Furthermore, for any $c\in \mathbb{R}$, we have that $\mathtt{M}_c:=\{(\mu_1',\mu_2'): \lambda(\mu_1)-\lambda(\mu_2)=c\}$
is the line of distributions within $\mathtt{M}^2$ with the same odds ratio $ \log (\mu_1(1-\mu_2) /((1-\mu_1)\mu_2))=c$.
The sequential probability ratio test of two proportions from \cite{Wald1947} was based on fixing a $c$ for the alternative (viewing it as a notion of `effect size') and analyzing sequences of paired data $X_{(1)}, X_{(2)}, \ldots$ with $X_{(i)} = (X_{i,1}, X_{i,2})$ 
by the product of conditional probabilities 
\[\frac{p_c(X_{(i)}\mid Z_{(i)})}{p_0(X_{(i)}\mid Z_{(i)})} = \Scond(X_i),\] thus effectively using $\Scond$ (here, we abuse notation slightly, writing $p_c(x\mid z) $ when we mean $p_{\mu'_1,\mu'_2}(x\mid z)$ for any $\mu'_1,\mu'_2\in \mathtt{M}_c$). It is, however, important to note that this product was not used for an anytime-valid test but rather for a classical sequential test with a fixed stopping rule especially designed to optimize power. 
\end{example}

\section{Growth Rate Comparison of Our E-variables}\label{Theoretical_results}
Above we provided several recipes for constructing \E-variables $S = S^{ \vec{\mu} }$ whose definition implicitly depended on the chosen alternative $\vec{\mu}$. To compare these, 
we  define, for any non-negative random variables $S_1^{\vec{\mu}}$ and $S_2^{\vec{\mu}}$, 
$S_1^{\vec{\mu}} \succeq S_2^{\vec{\mu}}$ to mean that for all $\vec{\mu} \in \mathtt{M}^k$, it holds that $\mathbb{E}_{P_{\vec{\mu}}}[\log S_1^{\vec{\mu}} ] \geq  \mathbb{E}_{P_{\vec{\mu}}}[\log S_2^{\vec{\mu}}]$.
We write $S^{\vec{\mu}}_1 \succ S^{\vec{\mu}}_2$ if $S^{\vec{\mu}}_1 \succeq S_2$ and there exists $\vec{\mu}\in \mathtt{M}^k$ for which equality does not hold.
From now on we suppress the dependency on $\vec{\mu}$ again, i.e. we write $S$ instead of $S^{\vec{\mu}}$.
We trivially have, for every underlying exponential family $\cM$, 
\begin{equation}\label{eq:simplerelation}
\Spseudo \succeq \Sripr \succeq \Smix \ \text{and}\ \Sripr \succeq \Scond.
\end{equation}
We proceed with Theorem~\ref{Taylor-approximation} and \ref{Taylor-approximation Scond} below (proofs in the Appendix). 
These results go beyond the qualitative assessment above, by numerically bounding  the difference in growth rate between $\Spseudo$ and $\Smix$ (and, because $\Sripr$ must lie in between them, also between these two and $\Sripr$) and $\Spseudo$ and $\Scond$ respectively.  Theorem~\ref{Taylor-approximation} and \ref{Taylor-approximation Scond} are asymptotic (in terms of difference between mean-value parameters) in nature.
To give more precise statements rather than asymptotics we need to distinguish between individual exponential families; this is done in the next section. 

To state the theorems, we need a notion of effect size, or discrepancy between the null and the alternative.
So far, we have taken the alternative to be fixed and given by $\vec{\mu}$, but effect sizes are usually defined with the null hypothesis as starting point. 
To this end, note that each $P_{\repvec{\mu_0}}\in \cH_0(\cM)$ corresponds to a whole set of alternatives for which $P_{\repvec{\mu_0}}$ is the closest point in KL within the null.
This set of alternatives is parameterized by $\mathtt{M}^{(k)}(\mu_0) = \{\mu'_1,\dots,\mu'_k\in \mathtt{M}: \frac1k \sum_{i=1}^k \mu'_i=\mu_0\}$, as in~\eqref{eq:mu0star}. 
We can re-parameterize this set as follows, using the special notation $\repvec{\mu_0}$ as given by (\ref{eq:repvec}). 
Let $\mathtt{A}$ be the set of unit vectors in ${\mathbb R}^k$ whose entries sum to 0, i.e. $ \vec{\alpha}\in \mathtt{A}$ iff  $\sqrt{\sum_{j=1}^k \alpha^2_j} = 1$ and $\sum_{j=1}^k \alpha_j = 0$. Clearly $\vec{\mu} \in \mathtt{M}^{(k)}(\mu_0)$ if and only if $\mu_1,\ldots,\mu_k \in \mathtt{M}$ and  $\vec{\mu} = \repvec{\mu_0} + \delta \vec{\alpha}$ for some scalar $\delta \geq 0$ and  $\vec{\alpha} \in \mathtt{A}$. 
We can think of $\delta$ as expressing the magnitude of an effect  and $\vec{\alpha}$ as its direction. Note that, if $k=2$, then there are only two directions, $\mathtt{A} = \{\vec{a}_{1},\vec{a}_{-1} \}$ with $\vec{a}_{1} = (1/\sqrt{2},-1/\sqrt{2})$ 
and $\vec{a}_{-1} = - \vec{a}_{1}$, corresponding to positive and negative effects: 
we have $\mu_1 - \mu_2 = \sqrt{2} \cdot \delta$ if $\vec{\alpha} = \vec{a}_1$ and $\mu_1 - \mu_2 = -\sqrt{2} \cdot \delta$ if $\vec{\alpha} = \vec{a}_{-1}$, as illustrated later on in Figure~\ref{fig:conjectures}. Also note that, for general $k$, in the theorem below, we can simply interpret $\delta$ as the Euclidean distance between $\vec{\mu}$ and $\repvec{\mu_0}$.
\begin{theorem}\label{Taylor-approximation}
Fix some $\mu_0 \in \mathtt{M}$, some $\vec\alpha \in \mathtt{A}$ and let $\vec{\mu} = \repvec{\mu_0} + \delta \vec{\alpha}$ for $\delta \geq 0$ such that $\vec{\mu} \in \mathtt{M}^{(k)}(\mu_0)$. The difference in growth rate between $\Spseudo$ and $\Smix$ is given by 
\begin{align}
\mathbb{E}_{P_{\vec{\mu}}}\left[ \log \Spseudo - \log \Smix \right] = \frac{1}{8} \int_{x} \frac{\left(f_x''(0) \right)^2}{f_x(0)} d \rho(x)  \cdot \delta^4 + o \left(\delta^4 \right) = O\left(\delta^4 \right),
\end{align}
where   $f_x(\delta) = \sum_{i=1}^k p_{\mu_0 + \delta \alpha_i}(x) = \sum\limits_{i=1}^k p_{\mu_i}(x)$ and $f_x''$ is the second derivative of $f_x$, so that $f_x(0) = k p_{\mu_0}(x)$ and (with some calculation) 
$f_x''(0) = \frac{d^2}{d \mu^2} p_{\mu}(x) \mid_{\mu = \mu_0}$.
\end{theorem}
As is implicit in the $O(\cdot)$-notation, the expectation on the left is well-defined and finite and the integral in the middle equation is finite as well. The theorem implies that for general exponential families, $\Smix$ is surprisingly close $(O(\delta^4))$ to the optimal $\Sripr$ in the GRO sense, whenever the distance $\delta$ between $\cH_1$ and $\cH_0(\cM)$ is small.  This means that, whenever $\Sripr \neq \Spseudo$ (so $\Sripr$ is hard to compute and $\Spseudo$ is not an \E-variable), we might consider using $\Smix$ instead: it will be more robust (since it is an \E-variable for the much larger hypothesis $\cH_0(\textsc{iid})$) and it will only be slightly worse in terms of growth rate.

Theorem \ref{Taylor-approximation}
is remarkably similar to the next theorem, which involves $\Scond$ rather than $\Smix$.
To state it, we first set $X_k(x^{k-1},z) := z- \sum_{i=1}^{k-1} x_i$, and we denote the marginal distribution  of $Z= \sum_{i=1}^k X_i$ under $P_{\vec{\mu}}$ as $P_{\vec{\mu};[Z]}$, noting that its density $p_{\vec{\mu};[Z]}$ is given by
\begin{equation}\label{eq:marginalfun}
p_{\vec{\mu};[Z]}(z) =  \int_{\mathcal{C}(z)} p_{\vec{\mu}}\left({x}^{k-1}, x_k \right) d\rho({x}^{k-1}),
\end{equation} 
where $\rho$ is extended to the product measure of $\rho$ on $\cX^{k-1}$ and
\begin{equation}\label{eq:cz}
\mathcal{C}(z) := \left\{{x}^{k-1} \in {\cal X}^{k-1} : X_i(x^{k-1},z) \in {\cal X} \right\}.
\end{equation} 
\begin{theorem}\label{Taylor-approximation Scond}
Fix some $\mu_0 \in \mathtt{M}$, $\vec\alpha \in \mathtt{A}$ and let $\vec{\mu} = \repvec{\mu_0} + \delta \vec{\alpha}$ for $\delta \geq 0$ such that $\vec{\mu} \in \mathtt{M}^{(k)}(\mu_0)$.
The difference in growth rate between $\Spseudo$ and $\Scond$ is given by 
\begin{align}
\mathbb{E}_{P_{\vec{\mu}}}\left[ \log \Spseudo - \log \Scond \right]=\frac{1}{8} \int_{z} \frac{\left(g''_z(0) \right)^2}{g_z(0)} d \rho_{[Z]}(z) \cdot \delta^4 + o \left(\delta^4 \right) = O(\delta^4),
\end{align} 
where $g_z(\delta) := p_{\repvec{\mu_0} + \vec\alpha \delta;[Z]}(z)$ and $\rho_{[Z]}$ denotes the measure on $Z$ induced by the product measure of $\rho$ on $\cX^k$; an explicit expression for  $g''_z(0)$ is 
\[\int_{\mathcal{C}(z)} p_{\repvec{\mu_0} }\left(\dataValue  \right) \sum\limits_{j=1}^k  \left[I'(\mu_0)(x_j-\mu_0) - I(\mu_0) \right]d\rho({x}^{k-1}),\] where  $I(\mu)$ denotes the Fisher information for $\mu$ and $I'(\mu)$ is its first derivative. 
\end{theorem}
Again, the expectation on the left is well-defined and finite and the integral on the right is finite. 
Comparing Theorem~\ref{Taylor-approximation Scond} to Theorem~\ref{Taylor-approximation}, we see that $f_x(0)$, the sum of $k$ identical densities evaluated at $x$, is replaced by $g_z(0)$, the density of the sum of $k$ i.i.d. random variables evaluated at $z$. 
\begin{corollary}
With the definitions  as in the two theorems above, the growth-rate difference
$\mathbb{E}_{P_{\vec{\mu}}}\left[ \log \Scond - \log \Smix \right]$ can be written as
\begin{align}
\frac{1}{8} \left(\int_{x} \frac{\left(f_x''(0) \right)^2}{f_x(0)} d \rho(x) - \int_{z} \frac{\left(g_z''(0) \right)^2}{g_{z}(0)} d \rho_{[Z]}(z) \right) \cdot \delta^4 + o \left(\delta^4 \right) = O\left(\delta^4 \right).
\end{align}
\end{corollary}

\section{Growth Rate Comparison for Specific Exponential Families}
\label{sec:specific}
We will now establish more precise relations between the four (pseudo-) \E-variables in $k$-sample tests for several standard exponential families, namely those listed in Table~\ref{tab:ordering} and a few related ones, as listed at the end of this section.
For each family $\cM$ under consideration, we give proofs for which different \E-variables are the same, i.e. $S = S'$, where $S, S' \in \{\Sripr, \Scond, \Smix, \Spseudo \}$.
Whenever we can prove that $\Sripr \neq S$ for another \E-variable  $S \in \{\Scond,\Smix\}$, we can infer that  $\Sripr \succ S$  because $\Sripr$ is the GRO \E-variable for $\cH_0(\cM)$.
Whenever both $\Scond$ and $\Smix$ are not equal to $\Sripr$, we will investigate via simulation whether $\Smix \succ \Scond$ or vice versa --- our theoretical results do not extend to this case.  
All simulations are carried out for the case $k=2$ in the paper.
Theorem~\ref{Taylor-approximation} and Theorem~\ref{Taylor-approximation Scond} show that in the neighborhood of $\delta=0$ ($\mu_1, \ldots, \mu_k$ all close together), the difference $\mathbb{E}_{P_{\vec{\mu}}}[\log S - \log S']$ is  of order $\delta^4$ when  $S, S'\in \{\Sripr, \Spseudo,$ $ \Smix,\Scond\}$.  Hence  in the figures we will show $(\mathbb{E}_{P_{\vec{\mu}}}[\log S - \log S'])^{1/4}$,  since then we expect the distances to increase linearly as we move away from the diagonal, making the figures more informative. 

\begin{figure}[p]
    \centering
    \begin{subfigure}[t]{\textwidth}
        \centering
        \includegraphics[width=0.3\textwidth]{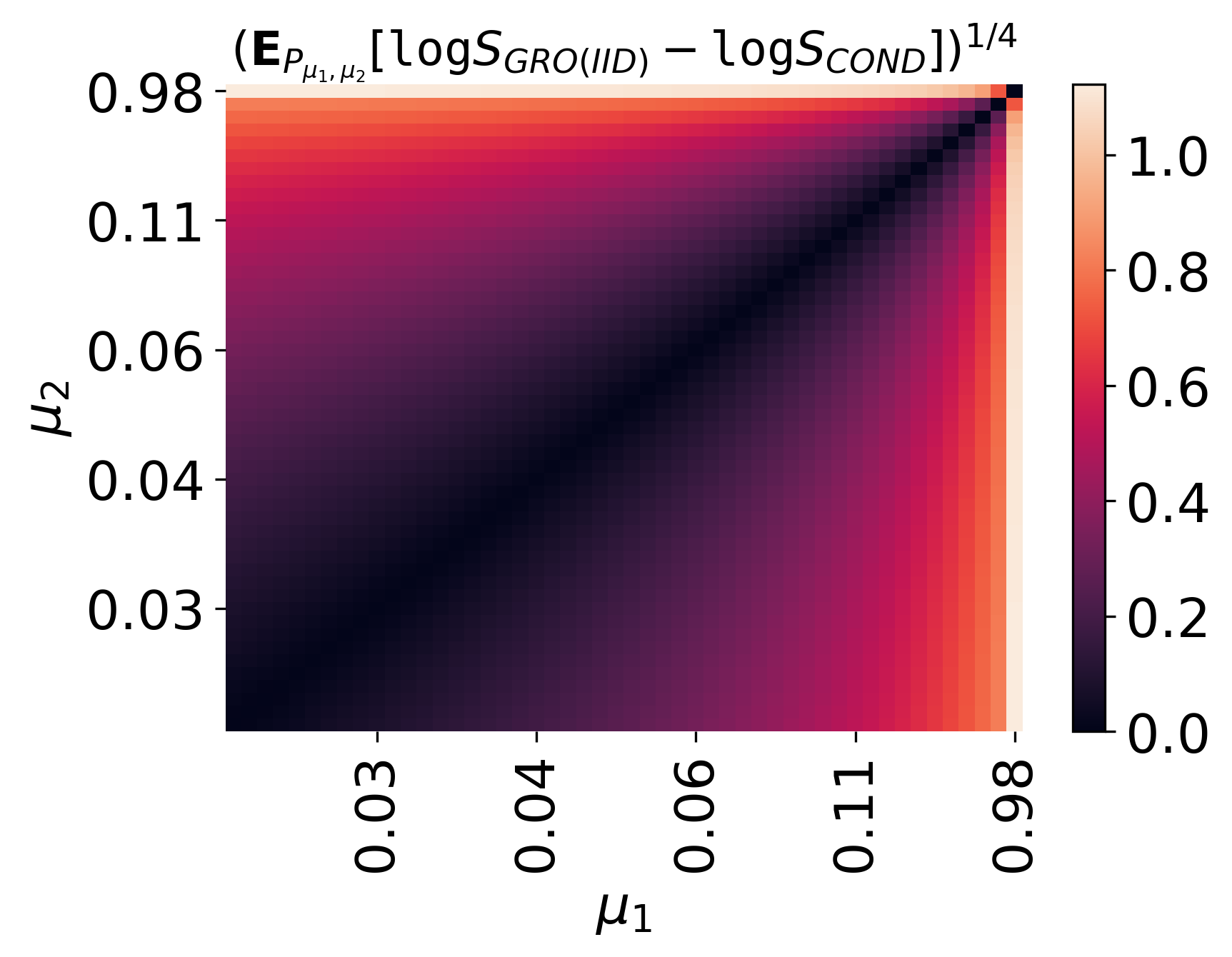}
        \includegraphics[width=0.335\textwidth]{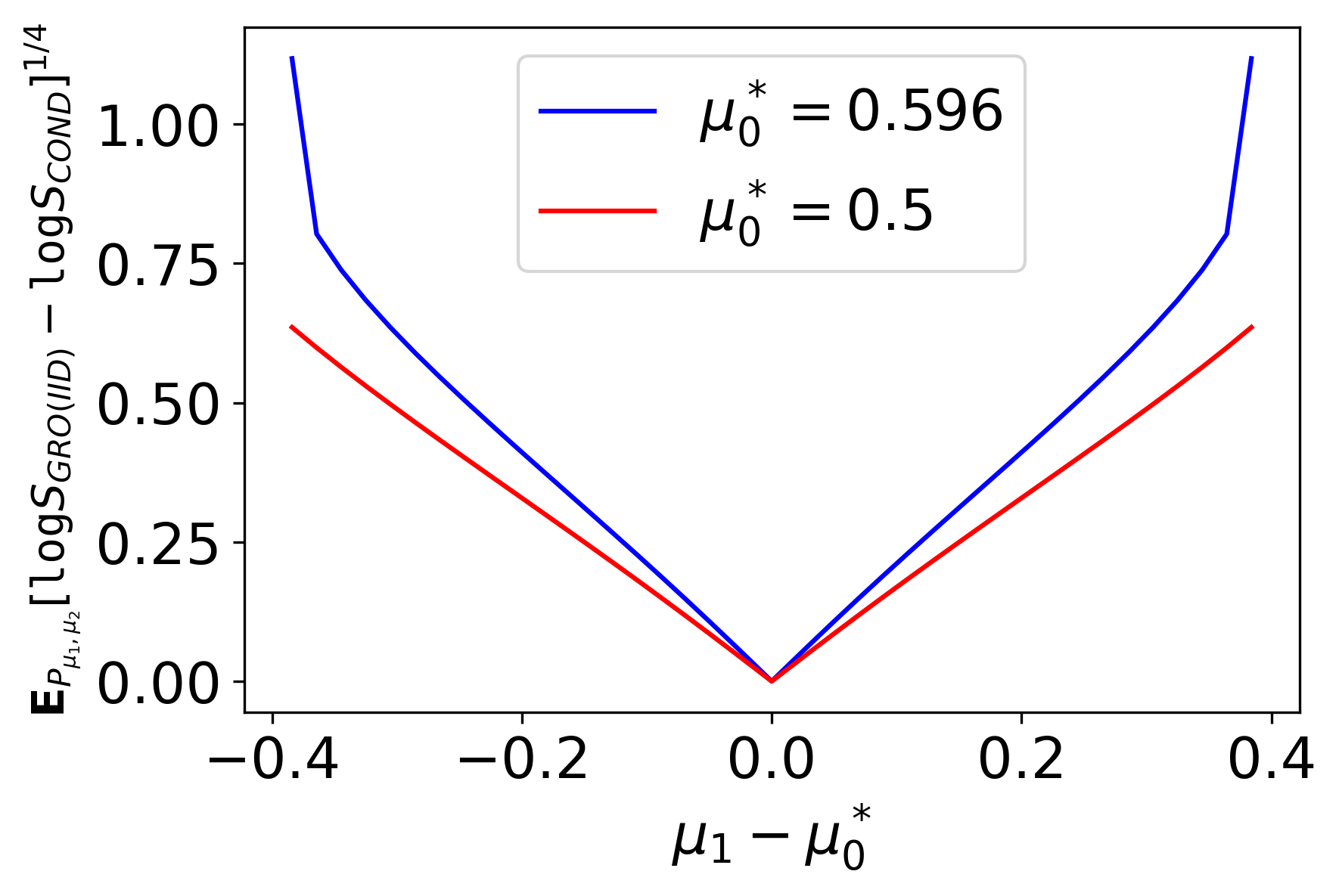}
        \includegraphics[width=0.33\textwidth]{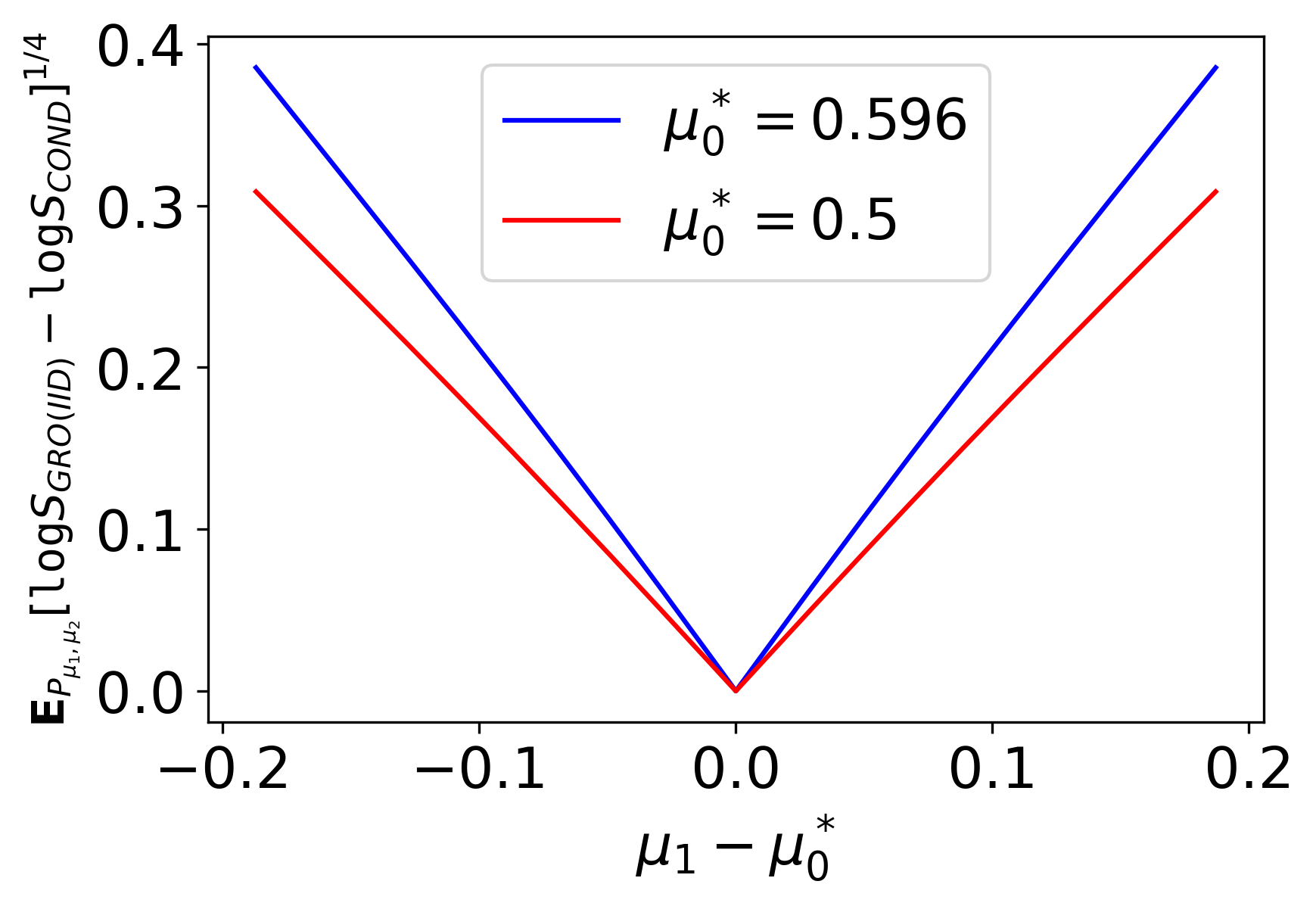}
        \caption{beta with free $\beta$ and fixed $\alpha$}
        \label{fig:conjecture_beta}
    \end{subfigure}
    \vspace{1em}
    \begin{subfigure}[t]{\textwidth}
        \centering
        \includegraphics[width=0.3\textwidth]{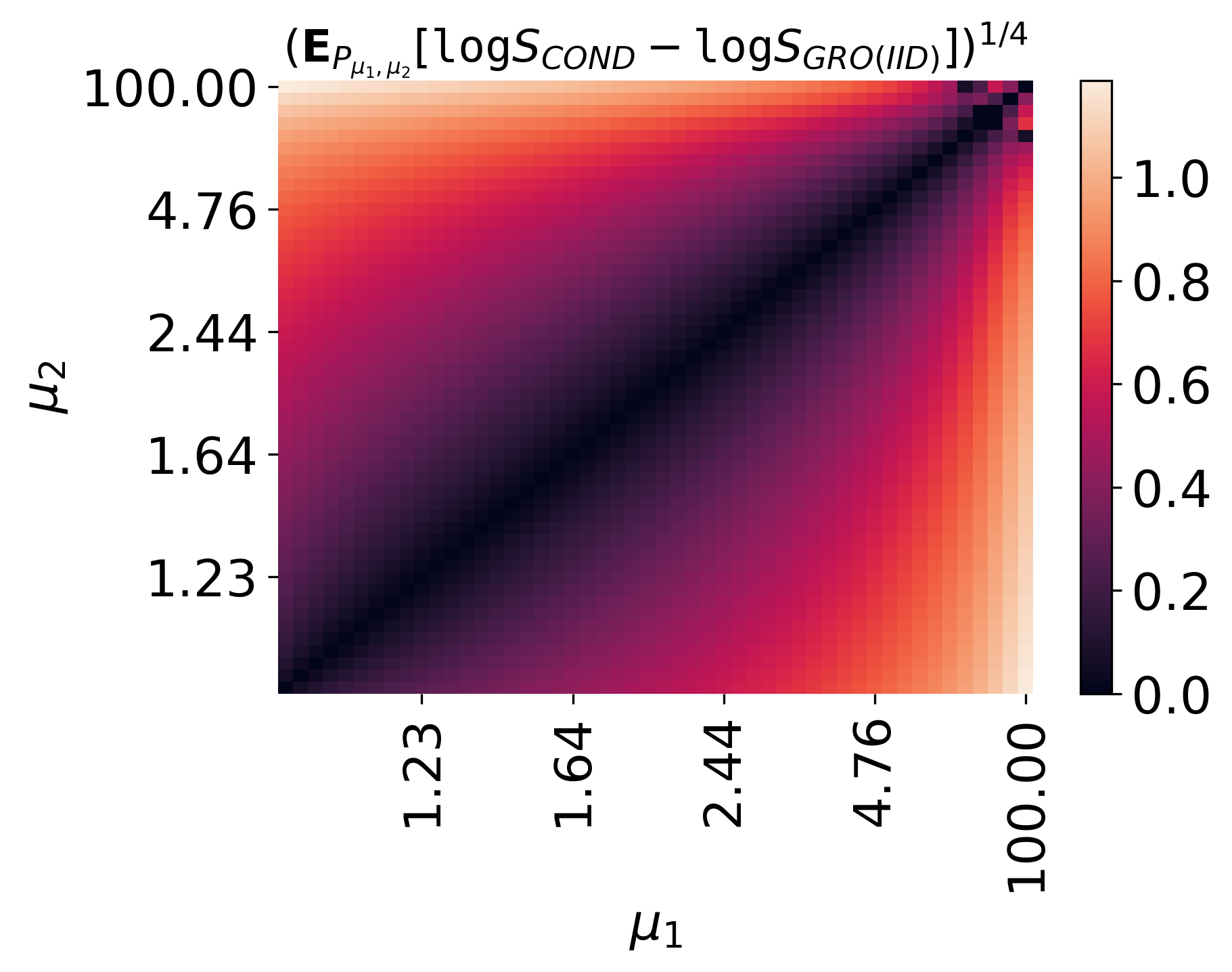}
        \includegraphics[width=0.33\textwidth]{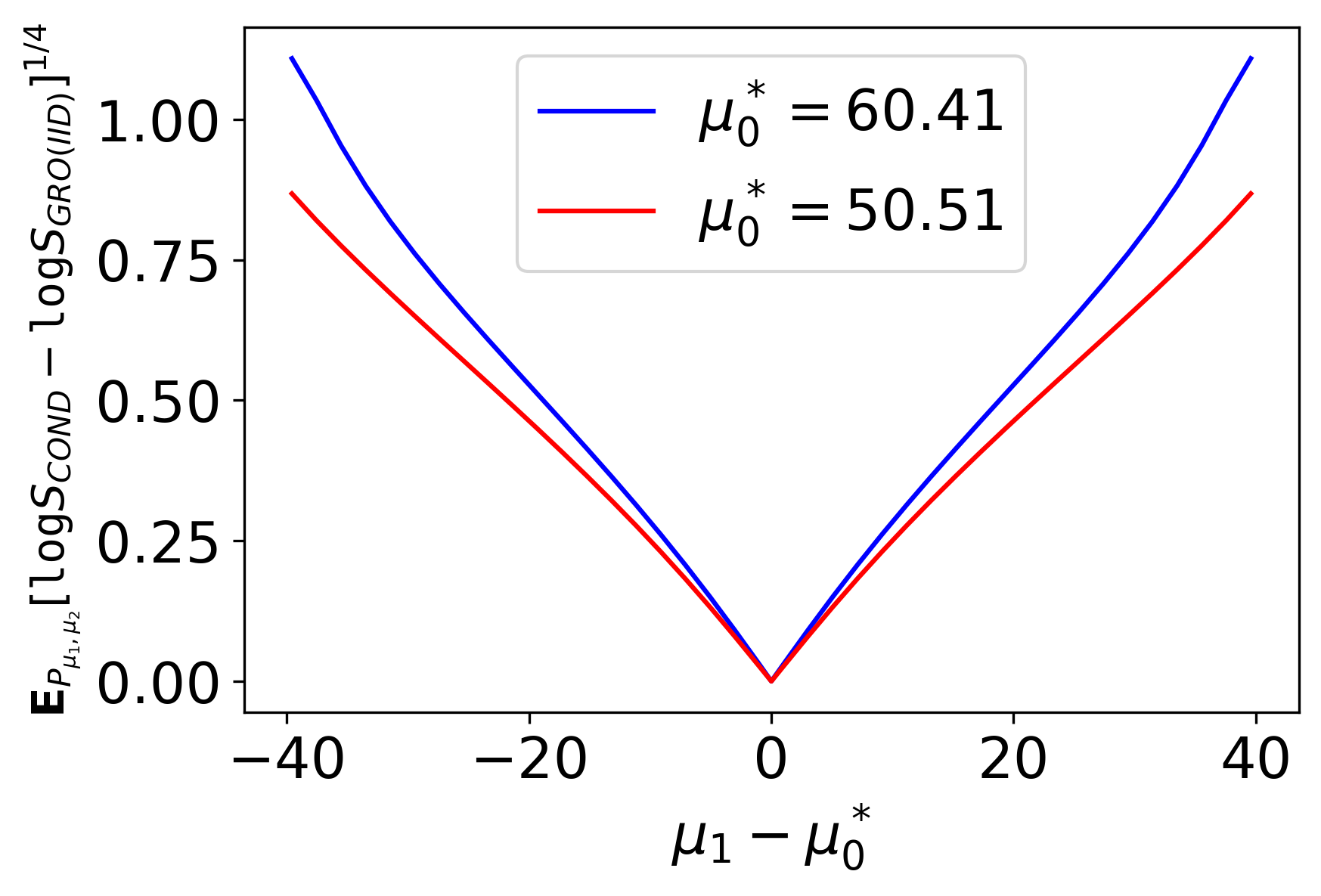}
        \includegraphics[width=0.335\textwidth]{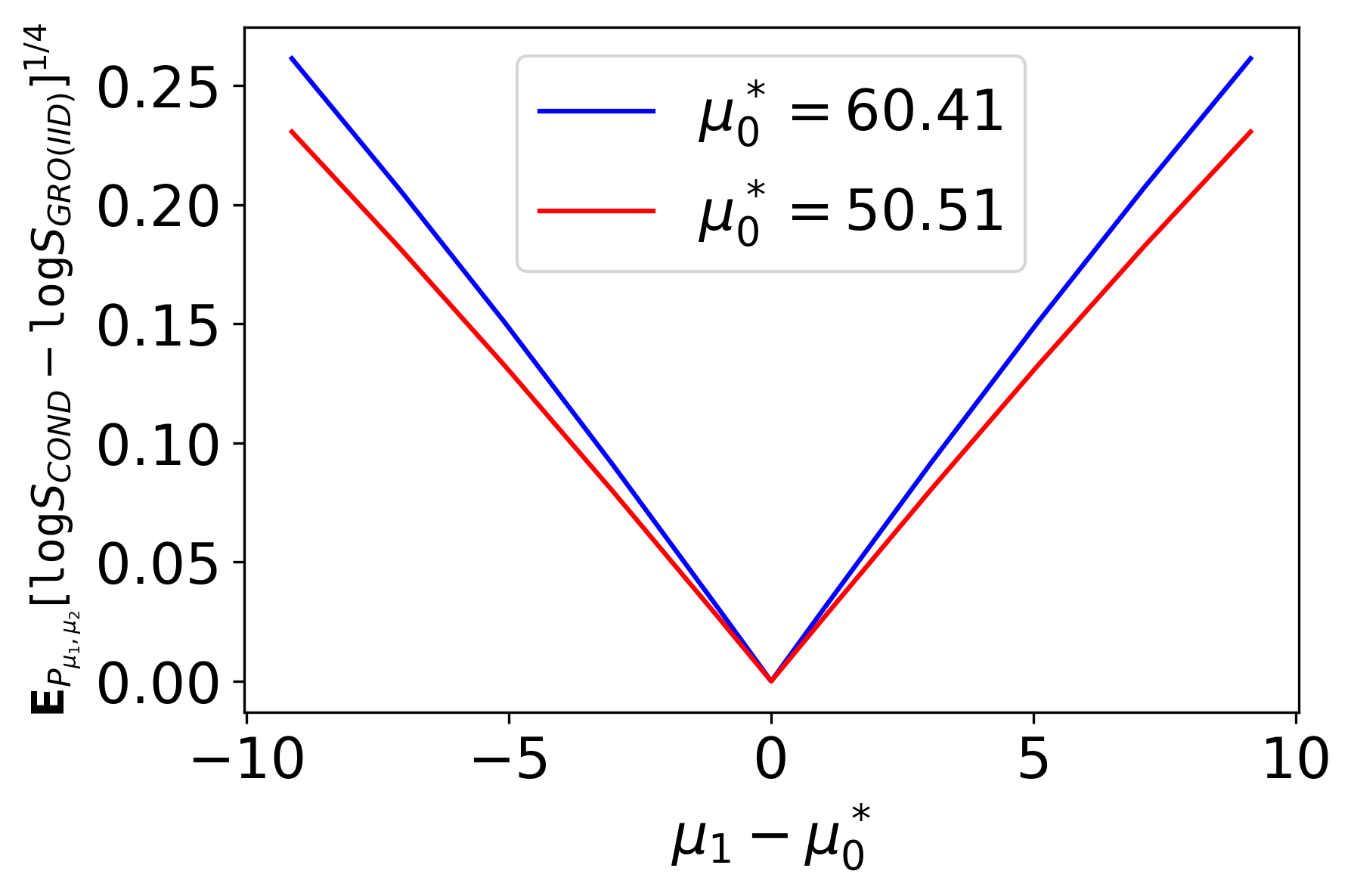}
        \caption{geometric}
        \label{fig:conjecture_geometric}
    \end{subfigure}
    \vspace{1em}
    \begin{subfigure}[t]{\textwidth}
        \centering
        \includegraphics[width=0.3\textwidth]{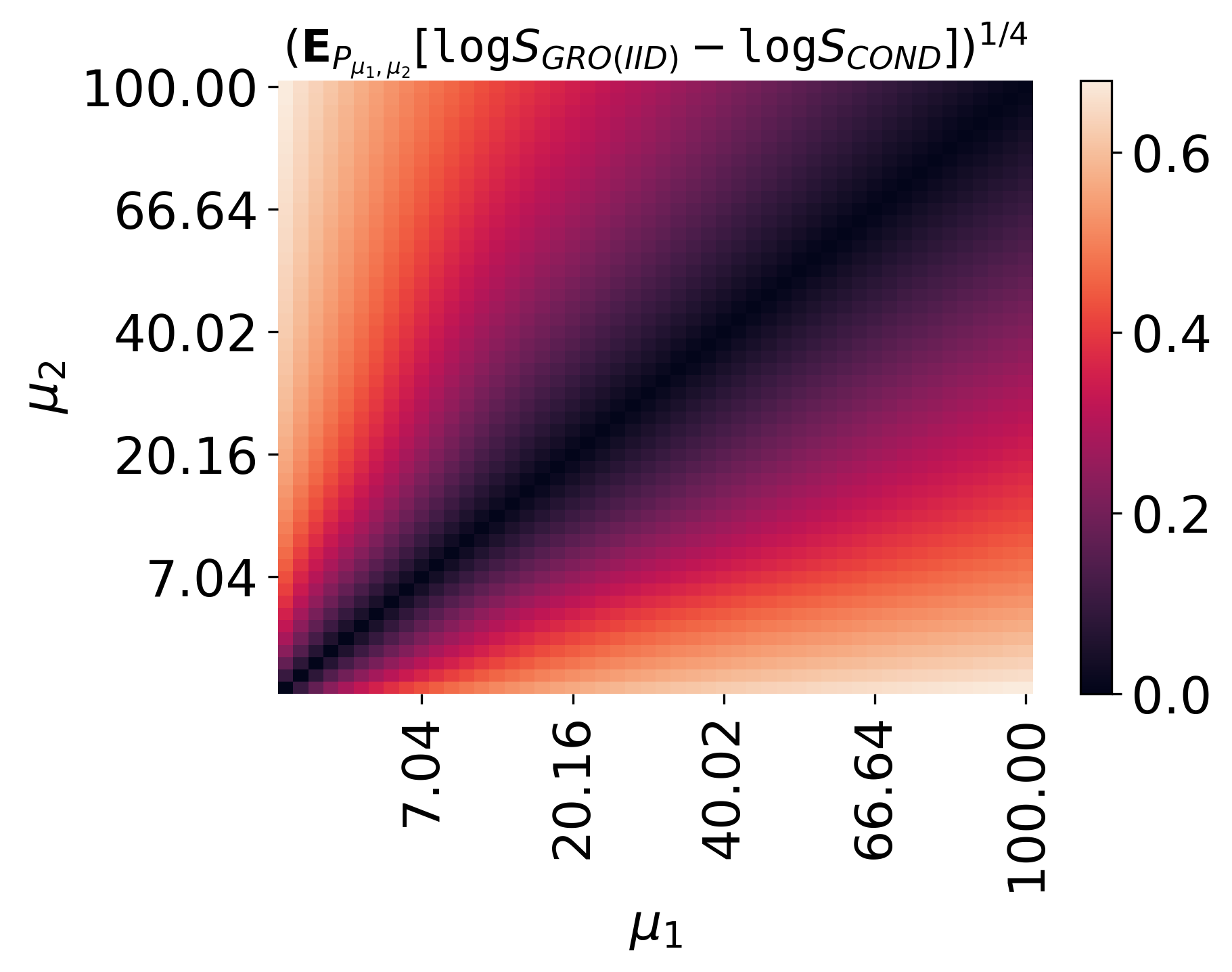}
        \includegraphics[width=0.33\textwidth]{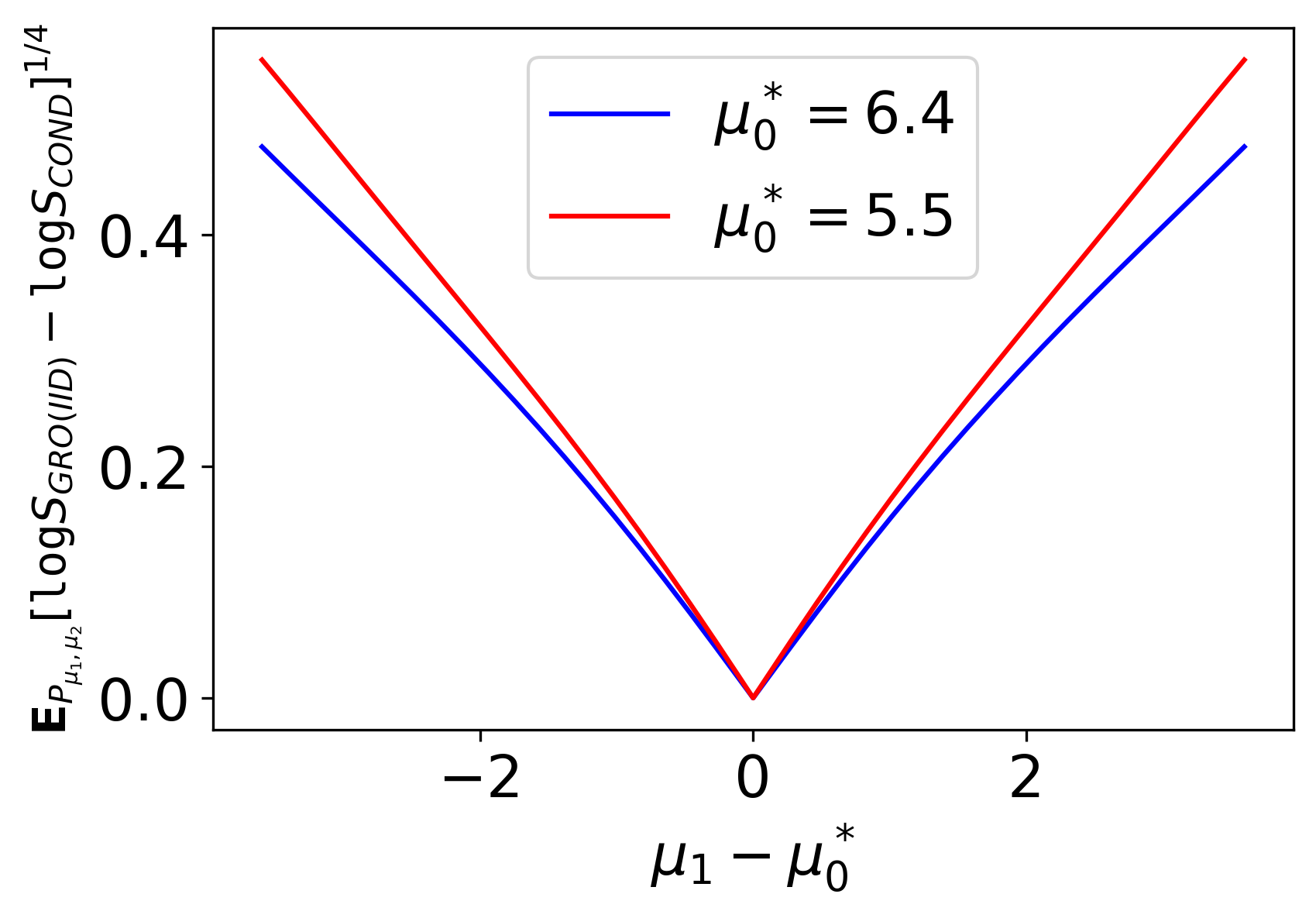}
        \includegraphics[width=0.33\textwidth]{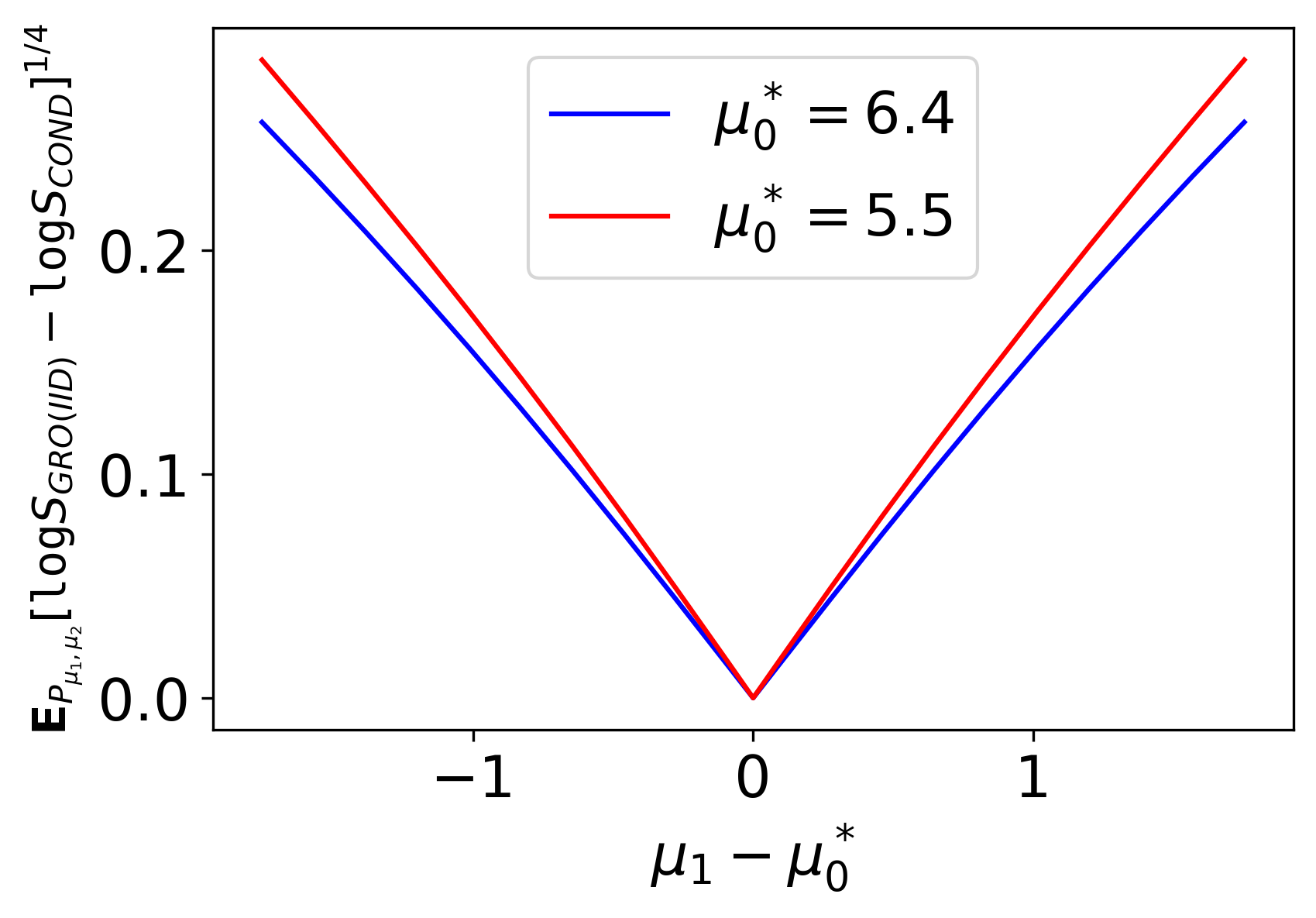}
        \caption{Gaussian with free variance and fixed mean}
        \label{fig:conjecture_gaussian}
    \end{subfigure}
    \vspace{1em}
    \begin{subfigure}[t]{\textwidth}
        \centering
        \includegraphics[width=0.3\textwidth]{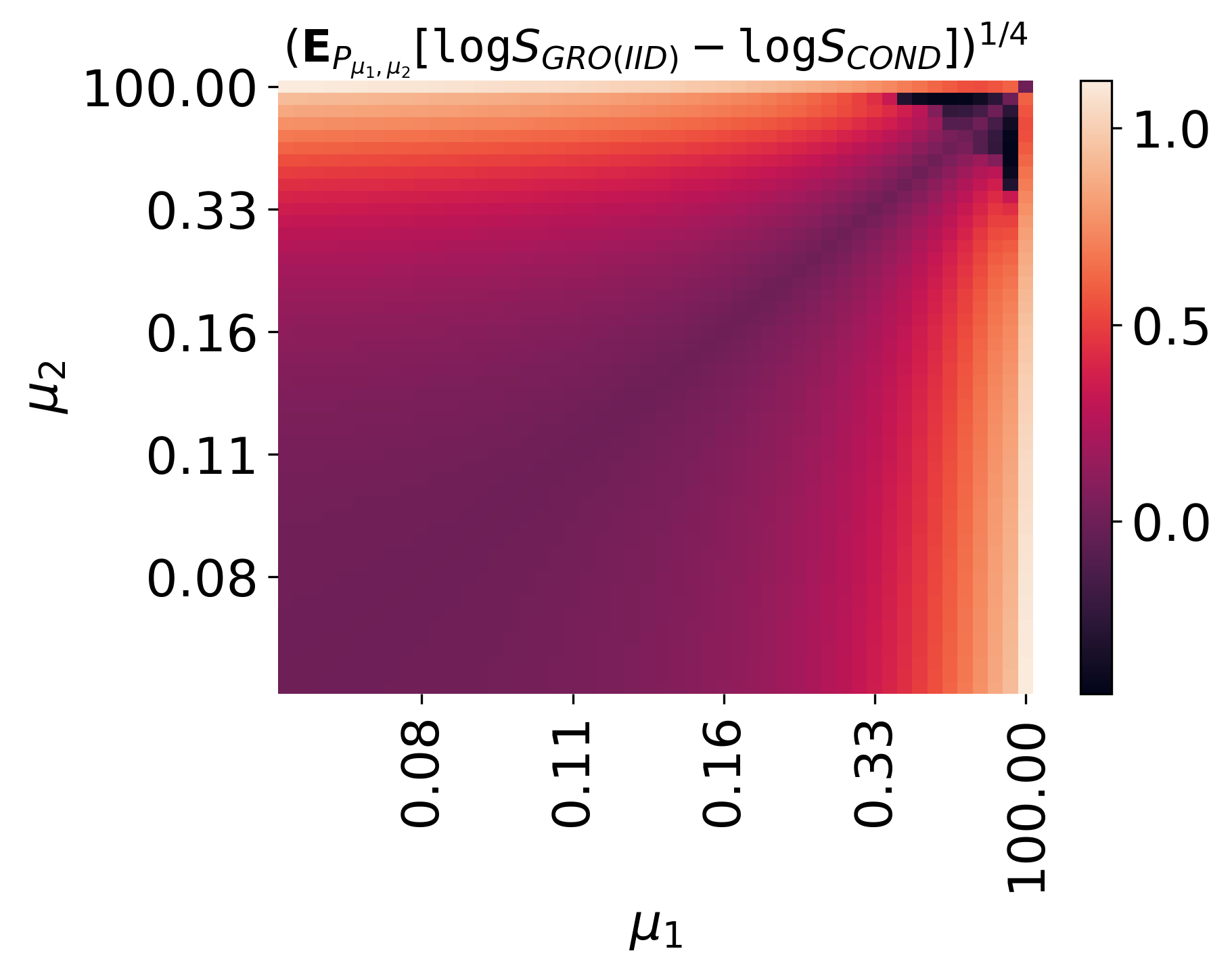}
        \includegraphics[width=0.32\textwidth]{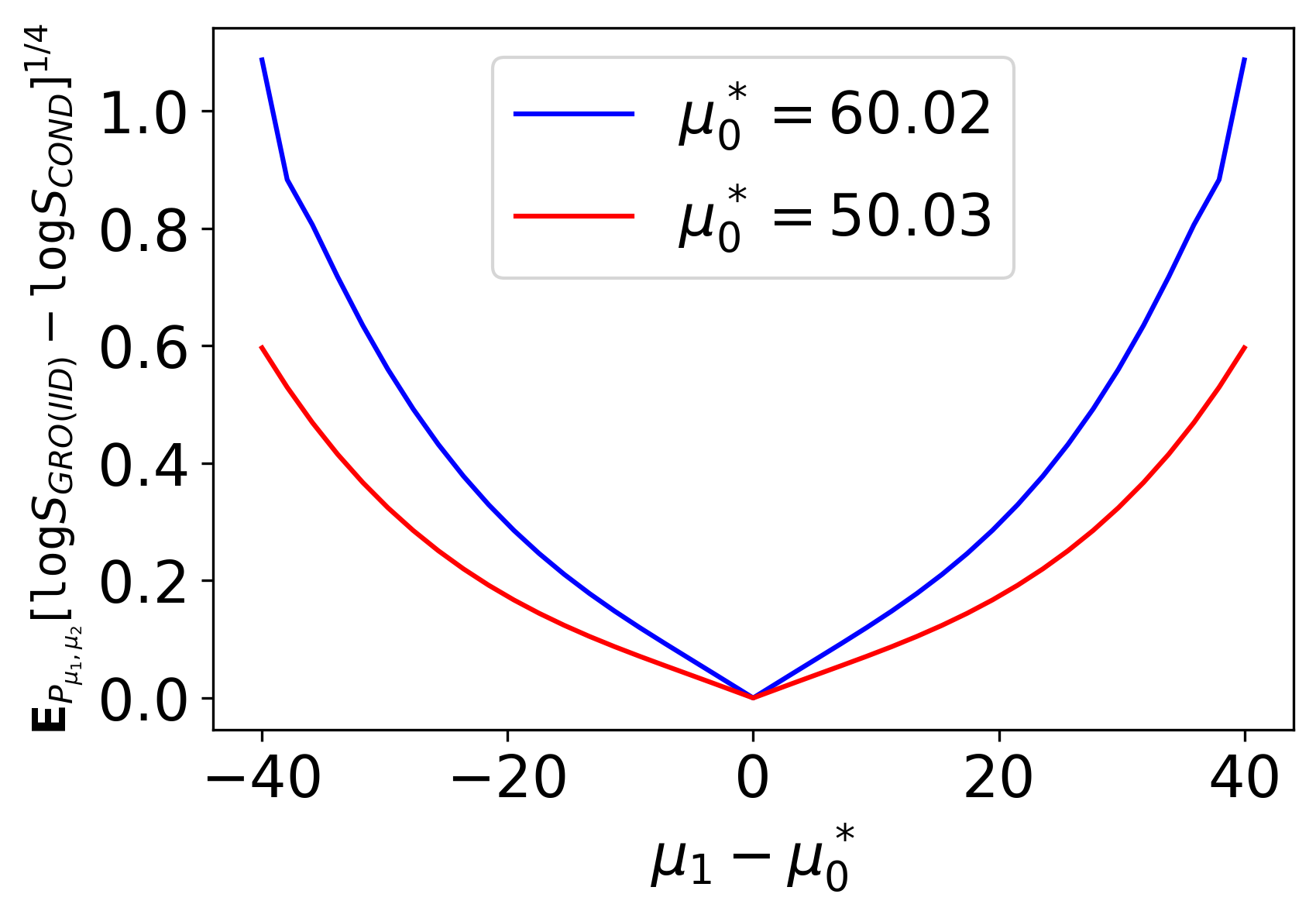}
        \includegraphics[width=0.34\textwidth]{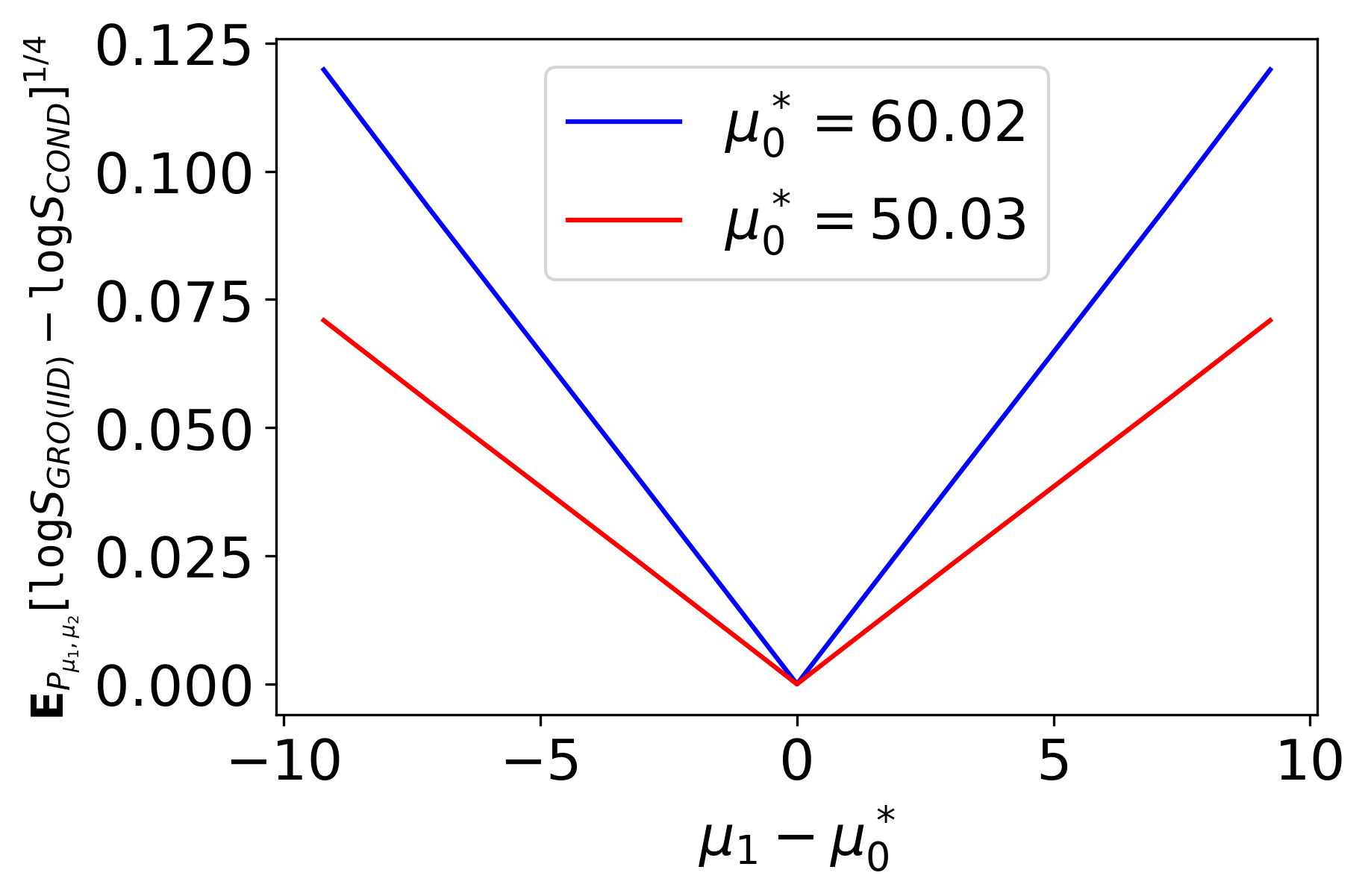}
        \caption{Exponential}
        \label{fig:conjecture_exponential}
    \end{subfigure}
    \caption{A comparison of $\Smix$ and $\Scond$ for four exponential families. We evaluated the expected growth difference on a grid of $50 \times 50$ alternatives $(\mu_1,\mu_2)$, equally spaced in the standard parameterization (explaining the nonlinear scaling on the depicted mean-value parameterization). On the left are the corresponding heatmaps. On the right are diagonal  `slices' of these heatmaps: the red curve corresponds to the main diagonal (top left - bottom right), the blue curve corresponds to the diagonal starting from the second tick mark (10th discretization point) top left until the second tick mark bottom right. These slices are symmetric around 0,  their value only depending on $\delta = \mid \mu_1 - \mu_2\mid /\sqrt{2} = \mid \mu_1 - \mu^*_0\mid \cdot \sqrt{2}$, where $\mu_0^* = (\mu_1 + \mu_2)/2$ and $\delta$ is as in Theorem~\ref{Taylor-approximation}
    }
    \label{fig:conjectures}
\end{figure}

Our findings, proofs as well as simulations, are summarised in Table~\ref{tab:ordering}.
For each exponential family, we list the rank of the (pseudo-)\E-variables when compared with the order `$\succ$'.
The ranks that are written in black are proven in Appendix~\ref{app:table_proofs}, while the ranks in blue are merely conjectures based on our simulations as stated above.
The results of the simulations on which these conjectures are based are given in Figure~\ref{fig:conjectures}.
Furthermore, the rank of $\Spseudo$ is colored red whenever it is not an \E-variable for that model, as shown in the Appendix. 
Note that whenever any of the \E-variables have the same rank, they must be equal $\rho$-almost everywhere, by strict concavity of the logarithm together with full support of the distributions in the exponential family. 
For example, the results in the table reflect that for the Bernoulli family, we have shown that $\Spseudo=\Sripr=\Smix$ and that $\Spseudo \succ \Scond$.
Also, for the geometric family and beta with free $\beta$ and fixed $\alpha$, we have proved that $\Spseudo$ is not an \E-variable, that $\Sripr \neq \Smix$ and that $\Sripr\neq \Scond$, so that it follows from~\eqref{eq:simplerelation} that $\Spseudo \succ \Sripr $, $\Sripr \succ \Smix$ and $\Sripr \succ \Scond$. 
Then the findings of the simulations shown in Figure~\ref{fig:conjecture_beta} suggest that $\Smix \succ \Scond$ for beta with free $\beta$ and fixed $\alpha$ and in Figure~\ref{fig:conjecture_geometric} suggest that $\Scond \succ \Smix$ for geometric family, but these are not proven. Figure~\ref{fig:conjecture_gaussian} shows that $\Smix \succ \Scond$ for Gaussians with free variance and fixed mean.
Finally, Figure~\ref{fig:conjecture_exponential} shows that for the exponential, there is no clear relation between $\Smix$ and $\Scond$. That is, $\Smix$ grows faster than $\Scond$ for some $\mu_1, \ldots, \mu_k \in \mathtt{M}$, and slower for others, which is indicated by rank $(3)-(4)$ in the table. 

\begin{table}[ht]
    \centering
    \begin{tabular}{lcccc}
    \toprule
        Exponential Family  & $\Spseudo$ & $\Sripr$ & $\Smix$ & $\Scond$  \\
         \midrule 
        Bernoulli & (1) & (1) & (1) & (2)\\
        Gaussian with free mean and fixed variance & (1) & (1) & (2) & (1)\\
        Poisson & (1) & (1) & (2) & (1)\\
        beta with free $\beta$ and fixed $\alpha$ & \textcolor{red}{(1)} & (2) & \textcolor{blue}{(3)} & \textcolor{blue}{(4)} \\
        geometric & \textcolor{red}{(1)} & (2) & \textcolor{blue}{(4)} & \textcolor{blue}{(3)} \\
        Gaussian with free variance and fixed mean & \textcolor{red}{(1)} & (2) & \textcolor{blue}{(3)} & \textcolor{blue}{(4)}\\
        Exponential & \textcolor{red}{(1)} & (2) & \textcolor{blue}{(3)-(4)} & \textcolor{blue}{(3)-(4)}\\
        \bottomrule
    \end{tabular}
    \caption{The ranks of the four different \E-variables when compared with the relation `$\succ$'. The ranks in black are proved in Appendix~\ref{app:table_proofs}, while the ranks in blue are conjectures based on the simulations in Figure~\ref{fig:conjectures}. The rank of $\Spseudo$ is denoted in red whenever it is not an \E-variable, as shown in Appendix~\ref{app:table_proofs}}
    \label{tab:ordering}
\end{table}

Finally, we note that for each family listed in the table, the results must extend to any other family that becomes identical to it if we reduce it to the natural form (\ref{ExponentialFormula}). For example,  the family of  Pareto distributions with fixed minimum parameter $v$ can be reduced to that of the  exponential distributions:  if  $U \sim \text{Pareto}(v, \alpha)$, then we can do a transformation $X = t(U)$ with $t(U)= \log(U/v)$, and  then $X \sim \text{Exp}(\alpha)$. Thus, the $k$-sample problem for $U$ with the Pareto$(v,\alpha)$ distributions, with $\alpha$ as free parameter,  is equivalent to the $k$-sample problem for $X$ with the exponential distributions; the \E-value $\Sripr$  obtained with a particular alternative in the Pareto setting for observation $U$ coincides with $\Sripr$ for the corresponding alternative in the exponential setting for observation $X=t(U)$, and the same holds for $\Smix$ and $\Scond$.
Therefore, the ordering for Pareto must be the same as the ordering for exponential in Table~\ref{tab:ordering}. 
Similarly, the \E-variables for the log-normal distributions (with free mean or variance) can be reduced to the two corresponding  normal distribution \E-variables. 
\section{Simulations to Approximate the RIPr}\label{sec:simulations} 
Because of its growth optimality property, we may sometimes still want to use  the GRO \E-variable $\Sripr$,
even in cases where it is not equal to the easily calculable $\Spseudo$. To this end we need to approximate it numerically. The goal of this section is twofold: first, we want to illustrate that this is feasible in principle; second, we show that this raises interesting additional questions for future work. 
Thus, below we consider in more detail simulations to approximate $\Sripr$ for the exponential families with $\Sripr \neq \Spseudo$ that we considered before, i.e. beta, geometric, exponential
and Gaussian with free variance; for simplicity we only consider the case $k=2$. In Appendix~\ref{simulations} we  provide some graphs illustrating the RIPr probability densities for particular choices of $\mu_1,\mu_2$; here, we focus on how to approximate them, taking our findings for $k=2$ as suggestive for what happens with larger $k$.

\subsection{Approximating the RIPr via Li's Algorithm}\label{Li_convergence}
\cite{li1999estimation} provides an algorithm for approximating the RIPr of distribution $Q$ with density $q$ onto  the convex hull $\textsc{conv}(\cP)$ of a set of distributions $\cP$ (where each $P \in \cP$ has density $p$) arbitrarily well in terms of KL divergence. 
At the $m$-th step, this algorithm outputs a finite mixture $P_{(m)}\in \textsc{conv}(\cP)$ of at most $m$ elements of  $\cP$. 
For $m>1$, these mixtures are determined by iteratively setting $P_{(m)} := \alpha P_{(m-1)} + (1-\alpha) P'$, where $\alpha\in [0,1]$ and $P'\in \cP$ are chosen so as to minimize KL divergence $D(Q \| \alpha {P}_{(m-1 )}+(1-\alpha)P')$.
Here, $P_{(1)}$ is defined as the single element of $\cP$ that minimizes $D(Q\|P_{(1)})$.
It is thus a greedy algorithm, but Li shows that, under some regularity conditions on $\cP$, it holds that $D(Q \| P_{(m)}) \rightarrow \inf_{P \in \textsc{conv}(\cP)} D(Q\|P)$. 
That is, $P_{(m)}$ approximates the RIPr in terms of KL divergence. 
This suggests, but is not in itself sufficient to prove, that $\sup_{P \in \cP} {\mathbb E}_P[q(X)/p_{(m)}(X)] \rightarrow 1$, i.e. that the likelihood ratio actually tends to an \E-variable. 

We numerically investigated whether this holds for our familiar setting with $k=2$, $Q$ is equal to $P_{\vec{\mu}}$ for some $\vec{\mu} = (\mu_1,\mu_2) \in {\mathtt M}^2$, and $\cP=\cH_0(\cM)$.
To this end, we applied Li's algorithm to a wide variety of values $(\mu_1,\mu_2)$ for the beta, exponential, geometric and Gaussian with free variance. 
In all these cases, after at most $m=15$ iterations, we found that
$\sup_{\mu_0\in \mathtt{M}} \mathbb{E}_{P_{\mu_0, \mu_0}}[p_{\mu_1,\mu_2}(X_1,X_2)/q_{(m)}(X_1,X_2)]$ 
was bounded by 1.005:  Li's algorithm convergences quite fast; see Appendix~\ref{simulations} for a graphical depiction of the convergence and design choices in the simulation. 

(note that, since we have proved that $\Sripr = \Spseudo$ for Bernoulli, Poisson and Gaussian with free mean, there is no need to approximate $\Sripr$ for those families).

\subsection{Approximating the RIPr via Brute Force}
\label{sec:brute}
While Li's algorithm converges quite fast, it is still highly suboptimal at iteration $m=2$, due to its being greedy.
This motivated us to investigate how `close' we can get to an e-variable by using a mixture of just two components. 
Thus, we set 
$p_{A}(x^k) := \alpha p_{\repvec{\mu_{01}}}(x^k) + (1-\alpha)p_{\repvec{\mu_{02}}} (x^k)$ and, for various choices of $\vec{\mu} = (\mu_1,\mu_2)$, 
considered 
\begin{equation}
\Sappr := \frac{p_{\vec{\mu}}(\data)}{p_{A}(\data)}
\end{equation}
as an approximate e-variable, for the specific values of  
$\alpha \in [0,1]$ 
and $\mu_{01}, \mu_{02}$ that minimize
$$\sup_{\mu_0\in \mathtt{M}} \mathbb{E}_{P_{\repvec{\mu_0}}}[\Sappr].$$
(in practice, we maximize $\mu_0$ over a discretization of $\mathtt{M}$ with $1000$ equally spaced grid points and minimize $\alpha, \mu_{01},\mu_{02}$ over a grid with $100$ equally sized grid points, with left- and right- end points of the grids over $\mathtt{M}$ determined by trial and error).

The simulation results, for $k=2$ and particular values of $\mu_1,\mu_2$ and the exponential families for which approximation makes sense (i.e. $\Sripr \neq \Spseudo$)  are presented in Table \ref{Table_optimal2Components_samePara}. We tried, and obtained similar results, for many more parameter values; one more parameter pair for each family is given in Table~\ref{Table_optimal2Components} in Appendix~\ref{simulations}. 
The term $\sup_{\mu_0 \in \mathtt{M}} \mathbb{E}_{P_{\repvec{\mu_0}}}[\Sappr]$ 
is remarkably close to 1 for all of these families. 
Corollary 2 of \citet{grunwald2019safe} implies that if the supremum {\em is\/} exactly 1, i.e. $\Sappr$ is an e-variable, then $\Sappr$ must also be the GRO e-variable relative to $P_{\vec{\mu}}$.
This leads us to speculate that perhaps all the exceedance beyond $1$ is due to discretization and numerical error, and the following might (or might not --- we found no way of either proving or disproving the claim) be the case:
\paragraph{Conjecture}
For $k=2$, the RIPr, i.e. the distribution achieving \[\min_{Q\in \textsc{conv}(\cH_0(\cM))}  D(P_{\mu_1,\mu_2}\|Q)\]
can be written as a mixture of just two elements of $\cH_0(\cM)$. 
\begin{table}
    \centering
    \begin{tabular}{ccccc}
    \toprule
    Distributions& $(\mu_1, \mu_2)$& $\alpha$& $(\mu_{01}, \mu_{02})$& $\sup\limits_{\mu_0 \in \mathtt{M}} \mathbb{E}_{X_1, X_2 \sim P_{\mu_0, \mu_0}}[S_{\textsc{appr}}]$\\
    \midrule
    beta&$(0.5, 0.25)$& $0.22$& (0.24, 0.81)& 1.0052\\
    Exponential& $(0.5, 0.25)$& 0.56& (0.35, 0.51)&  1.0000\\
    \makecell{Gaussian with free variance \\ and fixed mean} & $(0.5, 0.25)$&0.37 &(0.5, 0.5) &1.0000 \\
    \midrule
    Exponential &($\frac{10}{3}, \frac{5}{4}$)& 0.51& (0.62, 0.31)&  1.0047\\
    geometric&($\frac{10}{3}, \frac{5}{4}$)& 0.47& (1.84, 2.97)&  1.0008\\
    \makecell{Gaussian with free variance \\ and fixed mean} & $(\frac{10}{3}, \frac{5}{4})$&0.08 &(3.64, 2.73) & 1.0002 \\
    \bottomrule
    \end{tabular}
    \caption{For given values of $\vec{\mu} = (\mu_1,\mu_2)$, we show $\alpha,\mu_{01}$ and $\mu_{02}$ for the corresponding two-component mixture $\alpha p_{\mu_{01}}(X_1) p_{\mu_{01}}(X_2) + (1-\alpha) p_{\mu_{02}}(X_1) p_{\mu_{02}}(X_2)$ arrived at by brute-force minimization of the KL divergence as in Section~\ref{sec:brute}, and we show how close the corresponding likelihood ratio $S_{\textsc{appr}}$ is to being an e-variable  
    }
    \label{Table_optimal2Components_samePara}
\end{table}
\section{Conclusion and Future Work}
In this paper, we introduced and analysed four types of \E-variables for testing whether $k$ groups of data are distributed according to the same element of an exponential family.
These four \E-variables include: the GRO \E-variable ($\Sripr$), a conditional \E-variable ($\Scond$), a mixture \E-variable ($\Smix$), and a pseudo-\E-variable ($\Spseudo$).
We compared the growth rate of the \E-variables under a simple alternative where each of the $k$ groups has a different, but fixed, distribution in the same exponential family.  
We have shown that for any two of the \E-variables $S,S'\in \{\Sripr,\Scond,\Smix,\Spseudo\}$, we have $\mathbb{E}[\log S-\log S']=O(\delta^4)$ if the $\ell_2$ distance between the parameters of this alternative distribution and the parameter space of the null is given by $\delta$.
This shows that when the effect size is small, all the \E-variables behave surprisingly similar. 
For more general effect sizes, we know that $\Sripr$ has the highest growth rate by definition.
Calculating $\Sripr$ involves computing the reverse information projection of the alternative on the null, which is generally a hard problem.
However, we proved that there are exponential families for which one of the following holds $\Spseudo=\Sripr$, $\Scond=\Sripr$ or $\Smix=\Sripr$, which considerably simplifies the problem. 
If one is interested in testing an exponential family for which is not the case, there are algorithms to estimate the reverse information projection.
We have numerically verified that approximations of the reverse information projection also lead to approximations of $\Sripr$.
However, the use of $\Scond$ or $\Smix$ might still be preferred over $\Sripr$ due to the computational advantage.
Our simulations show that depends on the specific exponential family which of them is preferable over the other, and that sometimes there is even no clear order.

\paragraph{Acknowledgements} We thank Rosanne Turner and Wouter Koolen for various highly useful discussions. 
\paragraph{Declarations: financial support, lack of conflicting interests}
Partial financial support was received from China Scholarship Council State Scholarship Fund Nr.202006280045. 
The authors have no competing interests to declare that are relevant to the content of this article.
\bibliography{references}
\appendix
\section{Application in Practice: \texorpdfstring{$k$}{k} Separate I.I.D. Data Streams}
\label{sec:streaming}
In the simplest practical applications, we observe one block at a time, i.e. at time $n$, we have observed $\vec{X}_{(1)}, \ldots, \vec{X}_{(n)}$, where each $\vec{X}_{(i)} = (X_{i,1}, \ldots, X_{i,k})$ is a block, i.e. a vector with one outcome for each of the $k$ groups. 
This is a rather restrictive setup, but we can easily extend it to blocks of data in which each group has a different number of outcomes. For example, if data comes in blocks with $m_j$ outcomes in group $j$, for $j = 1\dots k$, $X_{(i)} = (X_{i,1,1}, \ldots, X_{i,1,m_1}, X_{i,2,1}, \ldots, X_{i,2,m_2}, \ldots, X_{i,k,1}, \ldots, X_{i,k,m_k})$, we can re-organize this having $k'= \sum_{j=1}^k m_j$ groups, having 1 outcome in each group, and having an alternative in which the first $m_1$ entries of the outcome vector share the same mean $\mu'_{1} = \ldots= \mu'_{m_1} = \mu_1$; the next $m_2$ entries share the same mean $\mu'_{m_1+1} = \ldots = \mu'_{m_1+m_2} = \mu_2$, and so on.

Even more generally though, we will be confronted with $k$ separate i.i.d streams and data in each stream may arrive at a different rate. 
We can still handle this case by pre-determining a multiplicity $m_1, \ldots, m_k$ for each stream. 
As data comes in, we fill virtual `blocks' with $m_j$ outcomes for group $j$, $j=1\dots k$.
Once a (number of) virtual block(s) has been filled entirely, the analysis can be performed as usual, restricted to the filled blocks.
That is, if for some integer $B$ we have observed $B m_j$ outcomes in stream $j$, for all $j=1\dots k$, but for some $j$, we have not yet observed  $(B+1) m_j$ outcomes, and we decide to stop the analysis and calculate the evidence against the null, then we output the product of \E-variables for the first $B$ blocks and ignore any additional data for the time being. 
Importantly, if we find out, while analyzing the streams, that some streams are providing data at a much faster rate than others, we may adapt $m_1, \ldots, m_k$ dynamically: whenever a virtual block has been finished, we may decide on alternative multiplicities for the next block; see \cite{turner2021safe} for a detailed description for the case that $k=2$.  
\section{Proofs for Section~\ref{sec:four}}
In the proofs we freely use, without specific mention, basic facts about derivatives of (log-) densities of exponential families. These can all be found in, for example, \cite{BarndorffNielsen78}.
\subsection{Proof of Proposition \ref{Speudo=Sripr}}
\begin{proof}
Since $\Sripr$ was already shown to be an E-variable in Lemma~\ref{lem:gro}, the `if' part of the statement holds. 
The `only-if' part follows directly from Corollary 2 to Theorem 1 in \citep{grunwald2019safe}, which states that there can be at most one E-variable of the form $p_{\vec{\mu}}(\data)/r(\data)$ where $r$ is a probability density for $\data$. 
\end{proof}
\subsection{Proof of Proposition \ref{SpseudoAnE-value}}
\begin{proof}
Define $g(\mu_0) := \mathbb{E}_{p_{\repvec{\mu_0}}}\left[\Spseudo \right]$ and $B(\mu_i) := A\left(\lambda(\mu_{i})+\lambda(\mu_{0})-\lambda(\mu_{0}^{*})\right)$.
\begin{align}\label{Expectaiton_Evalue}
g&(\mu_0) =
\mathbb{E}_{p_{\repvec{\mu_0}}}\left[\prod_{i=1}^{k} \frac{p_{\mu_{i}}\left(X_{i}\right)}{p_{\mu_{0}^{*}}\left(X_{i}\right)}\right]
=
\prod_{i=1}^{k} \mathbb{E}_{Y \sim p_{\mu_{0}}}\left[\frac{p_{\mu_{i}}(Y)}{p_{\mu_{0}^{*}}(Y)}\right]\nonumber \\
=&
\prod_{i=1}^{k} \int 
\exp \left(\lambda(\mu_{0}) y -A\left(\lambda(\mu_{0})\right)\right) \cdot \frac{
\exp \left(\lambda(\mu_{i}) y-A\left(\lambda(\mu_{i})\right)\right)}{
\exp \left(\lambda(\mu_{0}^{*}) y-A\left(\lambda(\mu_{0}^{*})\right)\right)}d 
\rho(y) \nonumber\\
=&
\prod_{i=1}^{k} \int 
\exp \left(\left(\lambda(\mu_{i})+\lambda(\mu_{0})-\lambda(\mu_{0}^{*})\right) y-A\left(\lambda(\mu_{i})\right)-A\left(\lambda(\mu_{0})\right)+A\left(\lambda(\mu_{0}^{*})\right)\right) d \rho(y) \nonumber \\
=&
\prod_{i=1}^{k} \exp \left(A\left(\lambda(\mu_{0}^{*})\right)-A\left(\lambda(\mu_{i})\right)-A\left(\lambda(\mu_{0})\right)\right) \exp \left(B(\mu_i) \right) \nonumber\\
& \qquad \cdot \int 
\exp \left(\left(\lambda(\mu_{i})+\lambda(\mu_{0})-\lambda(\mu_{0}^{*})\right) y - B(\mu_i) \right) d \rho(y) \nonumber \\
=&
\prod_{i=1}^{k} \exp 
\left(A\left(\lambda(\mu_{0}^{*})\right)-A\left(\lambda(\mu_{i})\right)-A\left(\lambda(\mu_{0})\right)\right) \exp \left(B(\mu_i)\right) \cdot 1\nonumber \\
=&
\exp \left(k A\left(\lambda(\mu_{0}^{*})\right)-\sum_{i=1}^{k} A\left(\lambda(\mu_{i})\right)-k A\left(\lambda(\mu_{0})\right)+\sum_{i=1}^{k} B(\mu_i) \right).
\end{align}
Taking first and second derivatives with respect to $\mu_0$, we find 
\begin{equation}\label{first derivative}
\frac{d}{d \mu_{0}} g(\mu_0)
=g(\mu_0) \cdot \frac{d}{d \mu_{0}}\left(\sum_{i=1}^{k} B(\mu_i) - k A\left(\lambda(\mu_{0})\right)\right)
\end{equation}
and
\begin{equation}
\begin{split}\label{Second_derivative}
\frac{d^2}{d \mu_{0}^2} g(\mu_0) 
=& \left( \frac{d}{d \mu_{0}} g(\mu_0) \right) \cdot \frac{d}{d \mu_{0}}\left(\sum_{i=1}^{k} B(\mu_i) - k A\left(\lambda(\mu_{0})\right)\right)
\\ &+ g(\mu_0) \cdot \frac{d^2}{d \mu_{0}^2}\left(\sum_{i=1}^{k} B(\mu_i) - k A\left(\lambda(\mu_{0})\right)\right)\\
=& g(\mu_0) \left(\sum\limits_{i=1}^k (\mu_i+\mu_0-\mu_0^{*}) - k\mu_0 \right)^2 
\\ &+ g(\mu_0) \left( \sum\limits_{i=1}^k \text{\sc var}_{P_{\mu_i+\mu_0-\mu_0^{*}}}[X] - k \text{\sc var}_{P_{\mu_0}}[X]\right)\\
=& g(\mu_0) \left( \sum\limits_{i=1}^k \text{\sc var}_{P_{\mu_i+\mu_0-\mu_0^{*}}}[X] - k \text{\sc var}_{P_{\mu_0}}[X]\right)
= g(\mu_0) \cdot f(\mu_0).
\end{split}
\end{equation}
where the second equality holds by (\ref{first derivative}), $( d/{d \lambda(\mu)} ) A(\lambda(\mu))= \mathbb{E}_{P_{\mu}}[X] $ and $({d^2}/{d \lambda(\mu)^2}) A(\lambda(\mu)) = \text{\sc var}_{P_{\mu}}[X]$.
(\ref{Second_derivative}) is continuous with respect to $\mu_0$. Therefore, if $f(\mu_0^{*})>0$ holds, it means that there exists an interval $\mathtt{M}^{*} \subset \mathtt{M}$ with $\mu_0^{*}$ in the interior of $\mathtt{M}^{*}$ on which  (\ref{Expectaiton_Evalue}) is strictly convex. Then there must exist a point $\mu_0' \in \mathtt{M}^{*}$ satisfying $\mathbb{E}_{P_{\repvec{\mu_{0}'}}}\left[\Spseudo \right] > \mathbb{E}_{P_{\repvec {\mu_{0}^*}}}\left[\Spseudo \right] = 1$, i.e. $\Spseudo$ is not an E-variable. Conversely, $f(\mu_0^{*})<0$ means that there exists an interval $\mathtt{M}^{*} \subset \mathtt{M}$ with $\mu_0^{*}$ in the interior of $\mathtt{M}^{*}$, on which (\ref{Expectaiton_Evalue}) is strictly concave. The result follows.
\end{proof}

\subsection{Proof of Theorem \ref{thm:smix}}
To prepare for the proof of Theorem~\ref{thm:smix}, let us first recall Young's [\citeyear{young1912classes}] inequality:
\begin{lemma}\label{Young's Inequality}{\bf [Young's inequality]}
Let $p, q$ be positive real numbers satisfying $\frac{1}{p} + \frac{1}{q} = 1$. Then if $a, b$ are nonnegative real numbers,
$
ab \leq \frac{a^p}{p} + \frac{b^q}{q}.
$
\end{lemma}
The proof of Theorem~\ref{thm:smix} follows exactly the same argument as the one used by \cite{turner2021safe} to prove this statement in the special case that $\cM$ is the Bernoulli model. 
\begin{proof}
We first show that $\Smix$ as defined in the theorem statement is  an E-variable. For this, we set $p^*_0(X) = \frac{1}{k} \sum\limits_{i=1}^k p_{\mu_i}(X)$. We have:
\begin{equation}\label{expectation_Smix}
\mathbb{E}_{\data \sim P_{\repvec{\mu_0}}}\left[\Smix \right]
= \mathbb{E}_{X_1 \sim P_{\mu_0}}\left[\frac{p_{\mu_1}(X_1)}{p^*_0(X_1)} \right] \cdot 
\ldots \cdot  \mathbb{E}_{X_k \sim P_{\mu_0}}\left[\frac{p_{\mu_k}(X_k)}{p^*_0(X_k)} \right].
\end{equation}
We also have
\begin{align}\label{condition_Smix}
& \frac{1}{k} \mathbb{E}_{X_1 \sim P_{\mu_0}}\left[\frac{p_{\mu_1}(X_1)}{p^*_0(X_1)} \right]
+ \cdots + \frac{1}{k} \mathbb{E}_{X_k \sim P_{\mu_0}}\left[\frac{p_{\mu_k}(X_k)}{p^*_0(X_k)} \right]\nonumber\\
=& \frac{1}{k} \mathbb{E}_{X \sim P_{\mu_0}}\left[
\frac{p_{\mu_1}(X)}{\frac{1}{k} \sum\limits_{i=1}^k p_{\mu_i}(X)} + \cdots + \frac{p_{\mu_k}(X)}{\frac{1}{k} \sum\limits_{i=1}^k p_{\mu_i}(X)}
\right]
= 1.
\end{align}
We need to show that (\ref{expectation_Smix}) $\leq 1$, for which we can use (\ref{condition_Smix}). Stated more simply, it is sufficient to  prove $\prod\limits_{i=1}^k r_i \leq 1$ with $\frac{1}{k}\sum\limits_{i=1}^k r_i \leq 1$, $r_i \in {\mathbb R}^+$. But this is easily established:
\begin{align}
\frac{1}{k}\sum\limits_{i=1}^k r_i 
= \frac{k-1}{k} \cdot \frac{\sum_{i=1}^{k-1} r_i}{k-1}
+ \frac{r_k}{k}
&\geq 
\left(\frac{\sum_{i=1}^{k-1} r_i}{k-1} \right)^{\frac{k-1}{k}} r_k^{\frac{1}{k}}\nonumber\\
&=
\left(\frac{k-2}{k-1}\cdot \frac{\sum_{i=1}^{k-2} r_i}{k-2} + \frac{r_{k-1}}{k-1} \right)^{\frac{k-1}{k}} r_k^{\frac{1}{k}}\nonumber\\
&\geq 
\left(\frac{\sum_{i=1}^{k-2} r_i}{k-2} \right)^{\frac{k-2}{k}} r_{k-1}^{\frac{1}{k}} r_k^{\frac{1}{k}}\nonumber\\
&\overset{\vdots}{\qquad} \nonumber\\
&\geq \left(\frac{r_1+r_2}{2} \right)^{\frac{2}{k}} \prod\limits_{i=3}^k r_i^{\frac{1}{k}} \geq \prod\limits_{i=1}^k r_i^{\frac{1}{k}}
\end{align}
where the first inequality holds because of Young's inequality, by setting $\frac{1}{p} := \frac{k-1}{k}, \frac{1}{q} := \frac{1}{k}, a^p := \frac{\sum_{i=1}^{k-1}r_i}{k-1}, b^q := r_k$ in Lemma \ref{Young's Inequality}. The other inequalities are established in the same way. It follows that  $\prod\limits_{i=1}^k r_i^{\frac{1}{k}} \leq 1$ and further $\prod\limits_{i=1}^k r_i \leq 1$.

This shows that $\Smix$ is a e-variable. 
It remains to show that $\Smix$ is indeed the GRO e-variable relative to $\cH_0(\textsc{iid})$; once we have shown this, it follows by Lemma 2 that it is the unique such e-variable and therefore by Lemma 1  that $P^*_0$ achieves the minimum in Lemma 1. 
Since we already know that $\Smix$ is an e-variable, the fact that it is the GRO e-variable relative to $\cH_0(\textsc{iid})$ follows immediately from Corollary 2 of Theorem 1 in  \cite{grunwald2019safe}, which states that there can be at most one e-variable of form $p_{\vec{\mu}}(\data)/r(\data)$ where $r$ is a probability density. Since $\Smix$ is such an e-variable, Lemma 1 gives that it must be the GRO e-variable.  
\end{proof}

\subsection{Proof of Proposition~\ref{prop:scond}}
\begin{proof}
The observed values of $X_1, X_2, \dots, X_{k}$ are denoted as $\dataValue$ (:= $x_1, \dots, x_k$).
With  $X_k(x^{k-1},z):= z - \sum_{i=1}^{k-1} x_i$ and  
$\mathcal{C}(z)$ as in (\ref{eq:cz}) and $ {p_{\vec{\mu}; [Z]} \left(z \right)}$ and $\rho(x^{k-1})$ as in (\ref{eq:marginalfun}), we get: 
\begin{align}
p&_{\vec{\mu}} \left({x}^{k-1} \middle \vert Z = z \right) 
= \frac{p_{\vec{\mu}} \left({x}^{k} \right)}
{p_{\vec{\mu}; [Z]} \left(z \right)}\nonumber\\
&= \frac{
\exp\left(\sum\limits_{i=1}^{k} \left( \lambda(\mu_i) x_i - A(\lambda(\mu_i))\right)\right)}{{ \displaystyle \int_{y^{k-1} \in \mathcal{C}(z)} }
\exp\left( \sum\limits_{i=1}^{k-1} \left(
\lambda(\mu_i) y_i - A(\lambda(\mu_i)) 
+ \lambda(\mu_k) X_k(y^{k-1},z))  - A(\lambda(\mu_k))
\right) \right)d\rho(y^{k-1})  } \nonumber \\
&= \frac{
\exp\left(\lambda(\mu_k)z+\sum\limits_{i=1}^{k-1} (\lambda(\mu_i)-\lambda(\mu_k)) x_i )\right)}{{ \displaystyle\int_{y^{k-1} \in \mathcal{C}(z)}}
\exp\left(\lambda(\mu_k)z+\sum\limits_{i=1}^{k-1}(\lambda(\mu_i)-\lambda(\mu_k)) y_i \right)
d\rho(y^{k-1}) 
}\nonumber \\
&=\frac{
\exp\left(\sum\limits_{i=1}^{k-1} (\lambda(\mu_i)-\lambda(\mu_k)) x_i\right)}{{ \displaystyle\int_{y^{k-1} \in \mathcal{C}(z)}}
\exp\left(\sum\limits_{i=1}^{k-1}(\lambda(\mu_i)-\lambda(\mu_k)) y_i\right)
d\rho(y^{k-1}) 
}\nonumber.
\end{align}
\end{proof}

\section{Proofs for Section~\ref{Theoretical_results}}
\subsection{Proof of Theorem \ref{Taylor-approximation}}
\begin{proof}
We prove the theorem using an elaborate Taylor expansion of $F(\delta)$, defined below, around $\delta =0$. We first calculate the first four derivatives of $F(\delta)$. Thus we define and derive, with $\mu_i = \mu_0 +\alpha_i\delta$ and  $f_y(\delta) = \sum\limits_{i=1}^k p_{\mu_i}(y)$ defined as in the theorem statement, 
\begin{align}\label{eq:above}
F(\delta) := & \mathbb{E}_{P_{\repvec{\mu_0}+ \vec{\alpha} \delta}}\left[\log \Spseudo - \log \Smix \right] \nonumber \\
=& 
\mathbb{E}_{P_{\vec{\mu}}}\left[ \log\prod\limits_{j=1}^k\left(\frac{1}{k}\sum\limits_{i=1}^k p_{\mu_i}(X_j) \right) 
-
\log p_{\repvec{\mu_0}}(\data) \right] \nonumber\\
=&
\mathbb{E}_{P_{\vec{\mu}}}\left[\sum\limits_{j=1}^k \log f_{X_j}(\delta) 
-
\sum\limits_{j=1}^k \log p_{\mu_0}(X_j) \right] - k\log k\nonumber\\
\overset{(a)}{=}&
\sum\limits_{j=1}^k \mathbb{E}_{X\sim P_{\mu_j}}\left[ \log f_{X}(\delta) 
-
\log p_{\mu_0}(X) \right] - k\log k\nonumber\\
\overset{(b)}{=} &
\overbrace{\int_{y \in {\cal X}} f_y(\delta)\log f_y(\delta)d \rho(y)}^{F_1(\delta)} + \overbrace{\left( - \int_{y \in {\cal X}} f_y(\delta)\log p_{\mu_0}(y)d \rho(y)\right)}^{F_2(\delta)} - k\log k,
\end{align}
where we define $F_1(\delta)$ to be equal to the leftmost term in (\ref{eq:above}) and $F_2(\delta)$ to be equal to the second, and $(a)$ and (b) both hold provided that
\begin{align}
 \label{eq:finitecheck}
\text{for all $j\in \{1,\ldots, k\}$:}\ 
\mathbb{E}_{X_j \sim P_{\mu_j}}
\left[ \mid \log f_{X_j}(\delta) -
 \log p_{\mu_0}(X_j) \mid  \right] < \infty
\end{align}
is finite. In the online supplementary material we verify that this condition, as well as a plethora of related finiteness-of-expectation-of-absolute-value conditions hold for all $\delta$ sufficiently close to $0$. Together these not just imply (a) and (b), but also (c) that we can freely exchange integration over $y$ and differentiation over $\delta$ for all such $\delta$  when computing the first $k$ derivatives of $F_1(\delta)$ and $F_2(\delta)$, for any finite $k$ and (d) that all these derivatives are finite for $\delta$ in a compact interval including $0$ (since the details are straightforward but quite tedious and long-winded we deferred these to the supplementary material). Thus, using (c), we will freely differentiate under the integral sign in the remainder of the proof below, and using (d), we will be able to conclude that the final result is finite.  

For each derivative, we first compute the derivative of $F_1(\delta)$ and then that of $F_2(\delta)$.
\begin{align}
F_1'(\delta)=& \int  f'_{y}(\delta)d \rho(y) + \int f'_{y}(\delta)\log f_{y}(\delta) d \rho(y) = 0,\nonumber\\
F_2'(\delta) =& -\int f_y'(\delta) \log p_{\mu_0}(y) d \rho(y) = 0, \text{\ so \ }
F'(0) = F_1'(0) + F_2'(0) = 0,
\end{align}
where the above formulas hold since $f_x'(0) = 0$ for all $x \in \cX$, which can be obtained by 
\begin{align}
f_x'(\delta^{\circ}) =& \sum\limits_{j=1}^k \frac{d p_{\mu_j}(x)}{d \mu_j}\frac{d \mu_j}{d \delta} (\delta^{\circ}) ,\nonumber\\
f_x'(0) =& \frac{d p_{\mu_0}(x)}{d \mu_0} \sum\limits_{j=1}^k \frac{d \mu_j}{d \delta}(0)
= \frac{d p_{\mu_0}(x)}{d \mu_0} \sum\limits_{j=1}^k \alpha_j
= 0,
\end{align}
where we used that all $\mu_j$ are equal to $\mu_0$ at $\delta = 0$. We turn to the second derivatives: 
\begin{align}
F_1''(\delta)=& \int f''_{y}(\delta)d \rho(y) + \int\left(f''_{y}(\delta)\log f_{y}(\delta) + \frac{\left(f'_y(\delta)\right)^2}{f_y(\delta)} \right) d \rho(y)  \nonumber \\ =& 
\int \left(f''_{y}(\delta)\log f_{y}(\delta) + \frac{\left(f'_y(\delta)\right)^2}{f_y(\delta)} \right) d \rho(y)\nonumber\\
F''_1(0) =&
\int \left(f''_{y}(0)\log f_{y}(0) + \frac{\left(f'_y(0)\right)^2}{f_y(0)} \right) d \rho(y);
 \nonumber \\  =&
\int f''_{y}(0)\log p_{\mu_0}(y) d \rho(y) + \int_{y \in {\cal X}}
\left( f''_{y}(0)\log k \right) d \rho(y)\\
=&
\int \left( f''_{y}(0)\log p_{\mu_0}(y) \right)  d \rho(y),\nonumber
\end{align}
where $\int f''_y(\delta)d \rho(y) = 0$ because $\int f_y(\delta)d \rho(y) = k$, in which $k$ is a constant that does not depend on $\delta$. 
Then $F_2''(\delta)$ is given by
\begin{align}
F_2''(\delta) =& -\int f_y''(\delta) \log p_{\mu_0}(y)d \rho(y)\ ;\ 
F_2''(0) = -\int f_y''(0) \log p_{\mu_0}(y)d \rho(y), \text{ so}\nonumber\\
F''(0) =& F_1''(0) + F_2''(0) = 0.
\end{align}
Now we compute the third derivative of $F(\delta)$, denoted as $F^{(3)}(\delta)$.
\begin{align}
F_1^{(3)}(\delta) =& \int  \left( f_y^{(3)}(\delta) \log f_y(\delta) + \frac{f_y''(\delta)f_y'(\delta)}{f_y(\delta)} + \frac{2f_y''(\delta)f_y'(\delta)f_y(\delta) - (f_y'(\delta))^3}{(f_y(\delta))^2} \right) d \rho(y) \nonumber\\
F_1^{(3)}(0) =& \int f_y^{(3)}(0) \log f_y(0) d \rho(y) \nonumber\\
=& \int f_y^{(3)}(0) \log p_{\mu_0}(y) d \rho(y) + \int f_y^{(3)}(0) \log k d \rho(y)\\ 
=& \int f_y^{(3)}(0) \log p_{\mu_0}(y) d \rho(y)\nonumber\\
F_2^{(3)}(\delta) =& -\int f_y^{(3)}(\delta) \log p_{\mu_0}(y) d \rho(y)\nonumber\\
F_2^{(3)}(0) =& -\int f_y^{(3)}(0) \log p_{\mu_0}(y) d \rho(y), \text{ so } F^{(3)}(0) = F_1^{(3)}(0) + F_2^{(3)}(0) = 0, \nonumber
\end{align}
which holds since $f_y'(0) = 0$ and $\int f_y(0) d \rho(y) = k$.\\~\\
The fourth derivative of $F(\delta)$ can be computed as follows:
\begin{align}
F_1^{(4)}(\delta)
=& \int \left( f_y^{(4)}(\delta) \log f_y(\delta) + \frac{f_y^{(3)}(\delta)f_y'(\delta)}{f_y(\delta)} \right) d \rho(y) \nonumber\\ 
&+ \int 3\cdot\frac{\left(f_y^{(3)}(\delta)f_y'(\delta)+(f_y''(\delta))^2\right)f_y(\delta) - f_y''(\delta)\left(f_y'(\delta)\right)^2}{\left(f_y(\delta)\right)^2} d \rho(y) \nonumber\\
&- \int \frac{3\left(f_y(\delta)f_y'(\delta)\right)^2\cdot f_y''(\delta) - 2\left(f_y'(\delta) \right)^4\cdot f_y(\delta)}{\left(f_y(\delta) \right)^4} d \rho(y) \ ; 
\end{align}
\begin{align}
F_1^{(4)}&(0)
= \int \left( f_y^{(4)}(0)\log f_y(0) + \frac{3\left(f_y''(0) \right)^2}{f_y(0)} \right) d \rho(y) \nonumber\\
=& \int f_y^{(4)}(0)\log p_{\mu_0}(y) d \rho(y) + \log k \int_{y \in {\cal X}} f_y^{(4)}(0) d \rho(y) + \int_{y \in {\cal X}} \frac{3\left(f_y''(0) \right)^2}{f_y(0)} d \rho(y) \nonumber\\
=& \int f_y^{(4)}(0)\log p_{\mu_0}(y) d \rho(y) + \int_{y \in {\cal X}} \frac{3\left(f_y''(0) \right)^2}{f_y(0)} d \rho(y),  \nonumber
\end{align}
and $F^{(4)}_2(\delta)$ can be computed by
\begin{align}
F_2^{(4)}(\delta)
=& -\int f_y^{(4)}(\delta)\log p_{\mu_0}(y) d \rho(y), \
F_2^{(4)}(0) = -\int f_y^{(4)}(0)\log p_{\mu_0}(y) d \rho(y), \text{ so} \nonumber\\
F^{(4)}(0) =& F^{(4)}_1(0) + F^{(4)}_2(0) = \int \frac{3\left(f_y''(0) \right)^2}{f_y(0)} d \rho(y) > 0. \nonumber
\end{align}
Based on the above derivatives, we can now do a fourth-order Taylor expansion of $F(\delta)$ around $\delta=0$, which gives: 
\begin{align*}
\mathbb{E}_{P_{\vec{\mu}}}\left[\log \Spseudo - \log \Smix \right]
=& \frac{1}{4!} F^{(4)}(0) \delta^4 + o(\delta^4) \\
=& \frac{1}{8} \int_{y \in {\cal X}} \frac{\left(f_y''(0) \right)^2}{f_y(0)} d \rho(y) \cdot \delta^4 + o \left(\delta^4 \right),
\end{align*}
where
$f_y(0)= \sum_{i=1}^k p_{\mu_0}(y) = k p_{\mu_0}(y)$ and 
$f_y''(0) = \left(\sum\limits_{i=1}^k \alpha_i^2 \right)\cdot \frac{d^2}{d \mu^2} p_{\mu}(y) \mid_{\mu = \mu_0} = \frac{d^2}{d \mu^2} p_{\mu}(y) \mid_{\mu = \mu_0}$.
\end{proof}

\subsection{Proof of Theorem \ref{Taylor-approximation Scond}}
\begin{proof}
We obtain the result using an even more involved Taylor expansion than in the previous theorem. As in that theorem, we will freely  differentiate (with respect to $\delta$) under the integral sign --- that this is allowed is again verified in the online supplementary material.  

Let $\vec{\mu}$, $\vec{\alpha}, \mathcal{C}(z),\rho(x^{k-1}), P_{\vec{\mu}}$ etc. 
be as in the theorem statement. We have: 
\begin{align}
f&(\delta) := \mathbb{E}_{P_{\vec{\mu}}}\left[\log \Spseudo - \log \Scond \right]\nonumber\\
=& 
\mathbb{E}_{P_{\vec{\mu}}}\left[\log \frac{p_{\vec{\mu}}\left(\data \right)}{p_{\repvec{\mu_0}}\left(\data \right)}
- \log \frac{p_{\vec{\mu}}\left({X}^{k-1} \mid 
Z 
\right)}{p_{\repvec{\mu_0}}\left({X}^{k-1} \mid
Z 
\right)} \right]\nonumber\\
=&  
\mathbb{E}_{P_{\vec{\mu}}}\left[\log \frac{p_{\vec{\mu}}\left(\data \right)}{p_{\repvec{\mu_0}}\left(\data \right)}
- \log \frac{p_{\vec{\mu}}\left(\data \right)}{p_{\repvec{\mu_0}}\left(\data \right)} 
+ \log \frac{\int_{\mathcal{C}(z)} p_{\vec{\mu}}\left(\dataValue \right) d\rho({x}^{k-1})}{\int_{\mathcal{C}(z)} p_{\repvec{\mu_0}}\left(\dataValue \right) d\rho({x}^{k-1})} \right] 
\nonumber \\
=&  
D\left( P_{\repvec{\mu_0} + \vec{\alpha} \delta; [Z]}\|P_{\repvec{\mu_0}; [Z]} \right)\nonumber.
\end{align}
We will prove the result by doing a Taylor expansion for $f(\delta)$  around $\delta = 0$. It is obvious that $f(0) = 0$ and the first derivative $f'(0) = 0$ since $f(0)$ is the minimum of $f(\delta)$ over an open set, and $f(\delta)$ is differentiable. We proceed to compute the second derivative of $f(\delta)$, 
using the notation $g_z(\delta) = p_{\repvec{\mu_0} + \vec\alpha \delta;[Z]}(z)$ as in the theorem statement, with $g'_z$ and $g''_z$ denoting first and second derivatives. 
\begin{align*}
f'(\delta)
=& 
\int 
g'_z(\delta) 
\log \frac{
g_z(\delta)} 
{g_z(0)} 
d\rho_{[Z]}(z) + \int 
g'_z(\delta) 
d\rho_{[Z]}(z) 
=
\int 
g'_z(\delta) 
\log \frac{
g_z(\delta)} 
{
g_z(0) 
} 
d\rho_{[Z]}(z). \\
f''(\delta)
=& \int 
g''_z(\delta) 
\log \frac{
g_z(\delta) 
}{
g_z(0) 
} 
d\rho_{[Z]}(z) 
+ \int 
\frac{\left(
g'_z(\delta) 
\right)^2}{
g_z(\delta) 
} 
d\rho_{[Z]}(z) 
,\\
\end{align*}
where in the first line, the second equality follows since the second term does not change if we interchanging differentiation and integration and the fact that $\int g_z(\delta) dz = 1$ is constant in $\delta$. We obtain 
\begin{equation}\label{eq:secondder}
f''(0) = \int 
\frac{\left(
g'_z(0) 
\right)^2}{
g_z(0) 
} 
d\rho_{[Z]}(z),
\end{equation}
and, with $x_k$ set to $X_k(x^{k-1},z)$ and recalling that $\vec{\mu} = \repvec{\mu_0} + \vec{\alpha} \delta$ and $\mu_j = \mu_0 + \alpha_j \delta$,
\begin{align*}
g'_z&(\delta) 
= \int_{\mathcal{C}(z)} \frac{d}{d \delta} 
p_{\repvec{\mu_0} + \vec{\alpha} \delta}
(x^k) d\rho({x}^{k-1})\\
=&
\int_{\mathcal{C}(z)} \sum\limits_{j=1}^k 
\prod_{i \in \{1,\ldots,k\} \setminus j} p_{\mu_i}(x_i)
\frac{d p_{\mu_j}(x_j)}{d \delta} d\rho({x}^{k-1})\\
=&
\int_{\mathcal{C}(z)} \sum\limits_{j=1}^k p_{\mu_1, \ldots, \mu_{j-1}, \mu_{j+1}, \ldots, \mu_k}(x_1, \ldots, x_{j-1}, x_{j+1}, \ldots, x_k) \frac{d p_{\mu_j}(x_j)}{d \mu_j} \frac{d \mu_j}{d \delta} d\rho({x}^{k-1})\\
=&
\int_{\mathcal{C}(z)} \sum\limits_{j=1}^k p_{\vec{\mu}}(x^k) \frac{d \log p_{\mu_j}(x_j)}{ d \mu_j} \alpha_j d\rho({x}^{k-1})\\
=&\int_{\mathcal{C}(z)} \sum\limits_{j=1}^k p_{\vec{\mu}}(x^k) \left(I(\mu_j)x_j - \mu_jI(\mu_j) \right) \alpha_j d\rho({x}^{k-1})
\end{align*}
where $I(\mu_j)$ is the  Fisher information. The final equality follows because, with $\lambda(\mu_j)$ the canonical parameter corresponding to $\mu_j$, we have $d \lambda(\mu_j)/d \mu_j = I(\mu_j)$ and  $dA(\beta)/d \beta)\mid_{\beta = \lambda(\mu_j)} = \mu_j$;   see e.g. \cite[Chapter 18]{grunwald2007minimum}. Now 
\begin{align}\label{eq:yundastrick}
g'_z(0) 
=&
\int_{\mathcal{C}(z)} \sum\limits_{j=1}^k p_{\repvec{\mu_0}}(x^k) \left(I(\mu_0)x_j - \mu_0 I(\mu_0) \right) \alpha_j d\rho({x}^{k-1})\nonumber \\
=&
\int_{\mathcal{C}(z)} p_{\repvec{\mu_0}}(x^k)I(\mu_0) \sum\limits_{j=1}^k x_j\alpha_j d\rho({x}^{k-1}) \\
=& I(\mu_0)
\cdot \int_{\mathcal{C}(z)} p_{\repvec{\mu_0}}(x^k)  \sum\limits_{j=1}^k x_j\alpha_j d\rho({x}^{k-1})
\end{align}
where the second equality follows from $\sum\limits_{j=1}^k \alpha_j = 0$.
Because $\data \ \text{\rm i.i.d.} \sim P_{\mu_0}$ under $P_{\repvec{\mu_0}}$ and the integral in (\ref{eq:yundastrick}) is over a set of exchangeable sequences, (For understanding the statement, we can consider the simple case $k=2$, $X_1$ and $X_2$ can be exchangeable because they are `symmetric' for given $\mathcal{C}(z)$.) we must have that  (\ref{eq:yundastrick}) remains valid if we re-order the $\alpha_j$'s in round-robin fashion, i.e. for all $i=1..k$, we have, with $\alpha_{j,i}= \alpha_{(j+i-1 )\mod k}$,
$$
g'_z(0)= I(\mu_0)
\cdot \int_{\mathcal{C}(z)} p_{\repvec{\mu_0}}(x^k)  \sum\limits_{j=1}^k x_j\alpha_{j,i} d\rho({x}^{k-1}).
$$
Summing these $k$ equations we get, using  that 
 $\sum\limits_{i=1}^k \alpha_i = 0$, that $k g'_z(0) = 0$  so that $g'_z(0) = 0$.
From (\ref{eq:secondder}) we now see that
$$
f''(0) = 0.
$$
Now we compute the third derivative of $f(\delta)$, denoted as $f^{(3)}(\delta)$:
\begin{align}
f^{(3)}(\delta)
=& \int 
\left( g^{(3)}_z(\delta)
\log \frac{
g_z(\delta) 
}{
g_z(0) 
} 
+ \frac{
g''_z(\delta) g'_z(\delta) 
}{
g_z(\delta) 
} \right) 
d\rho_{[Z]}(z) 
\nonumber\\ &
+ 
\int 
\left( \frac{2
g''_z(\delta) g'_z(\delta) g_z(\delta) - (g'_z(\delta))^3
}{
(g_z(\delta))^2
}
\right) d\rho_{[Z]}(z) 
\nonumber
\end{align}
So since $g'_z(0) = 0$ we must also have
$$
f^{(3)}(0) = 0.
$$
The fourth derivative of $f(\delta)$ is now computed as follows:
\begin{align}
f^{(4)}(\delta) =& \int 
\left( 
g^{(4)}_z(\delta) 
\log \frac{
g_z(\delta) 
}{
g_z(0)
}
+ \frac{
g^{(3)}_z(\delta) \cdot g'_z(\delta) 
}{
g_z(\delta) 
} \right) 
d\rho_{[Z]}(z) 
\nonumber\\ 
&+ \int 
3\cdot\frac{
\left(
g^{(3)}_z(\delta) \cdot g'_z(\delta) + (g''_z(\delta))^2
\right) 
g_{z}(\delta)
- 
g''_z(\delta) \cdot ( g'_z(\delta))^2
}
{
(g_z(\delta))^2
} 
d\rho_{[Z]}(z).
\nonumber
\end{align}
Then
\begin{align}
f^{(4)}(0) = 
\int 
\frac{3\left(
g''_z(0)
\right)^2}{
g_z(0) 
}
d\rho_{[Z]}(z) 
> 0.\nonumber
\end{align}
We now have all ingredients for a fourth-order Taylor expansion of $f(\delta)$ around $\delta =0$, which gives: 
\begin{align*}
\mathbb{E}_{P_{\vec{\mu}}}\left[\log \Spseudo - \log \Scond \right]
= \frac{1}{8} \int 
\frac{\left(g''_z(0)  \right)^2}{g_z(0) } 
d\rho_{[Z]}(z) 
\cdot \delta^4 + o \left(\delta^4 \right)
\end{align*}
which is what we had to prove.
\end{proof}

\section{Proofs for Section~\ref{sec:specific}}\label{app:table_proofs}
In this section, we prove all the statements in Table~\ref{tab:ordering}. 

\subsection{Bernoulli Family}
We prove that for $\cM$ equal to the Bernoulli family, we have $\Spseudo = \Sripr = \Smix \succ \Scond$. 
\begin{proof}
We set $\mu_0^* = \frac{1}{k} \sum\limits_{i=1}^k \mu_i$.
\begin{align}
\Smix 
:= \frac{p_{\vec{\mu}}(\data)}{\prod\limits_{j=1}^k \left(\frac{1}{k} \sum\limits_{i=1}^k p_{\mu_i}(X_j)\right)} 
&= \frac{p_{\vec{\mu}}(\data)}{\prod\limits_{j=1}^k \left(\frac{1}{k} \sum\limits_{i=1}^k \left(\mu_i^{X_j} (1-\mu_i)^{1-X_j} \right)\right)} \\ 
&= \frac{p_{\vec{\mu}}(\data)}{\prod\limits_{j=1}^k \left((\mu_0^*)^{X_j} (1-\mu_0^*)^{1-X_j} \right)}\nonumber\\
&= \frac{p_{\vec{\mu}}(\data)}{\prod\limits_{j=1}^k p_{\mu_0^*}(X_j)}
= \Spseudo
\end{align}
where the third equality holds since $X_i \in \{0, 1\}$. So $\Spseudo$ is an E-variable and $\Spseudo = \Sripr$ according to Theorem \ref{Speudo=Sripr}. Then the claim follows using (\ref{eq:simplerelation}) together with the fact that when $Z=0$ or $Z=2$, we have $\Scond=1$, while this is not true for the other \E-variables, so that $\Scond\neq \Sripr=\Spseudo=\Smix$. The result then follows from \eqref{eq:simplerelation}.
\end{proof}

\subsection{Poisson and Gaussian Family With Free Mean and Fixed Variance}\label{app:table_gaussfreemean}
We prove that for $\cM$ equal to the family of Gaussian distributions with free mean and fixed variance $\sigma^2$, we have $\Spseudo = \Sripr = \Scond \succ \Smix$.
The proof that the same holds for $\cM$ equal to the family of Poisson distributions is omitted, as it is completely analogous.
\begin{proof}
Note that if we let $Z:=\sum_{i=1}^k X_i$, then we have that $Z\sim \mathcal{N}(\sum_{i=1}^k \mu_i, k\sigma^2)$ if $X^k\sim P_{\vec{\mu}}$. 
Let $\mu^*_0$ be given by (\ref{eq:mu0star}) relative to fixed alternative $P_{\vec{\mu}}$ as in the definition of $\Spseudo$ underneath (\ref{eq:mu0star}).
Since $k {\mu^*_0}=\sum_{i=1}^k \mu_i$, we have that $Z$ has the same distribution for $\data \sim P_{\repvec{\mu^*_0}}$.
This can be used to write
\[\Scond = \frac{p_{\vec{\mu}}\left(\data\mid Z\right)}{p_{\repvec{\mu^*_0}}\left(\data\mid Z\right)}=
\frac{p_{\vec{\mu}}\left(\data\right)}{p_{\repvec{\mu^*_0}}\left(\data\right)}\frac{p_{\repvec{\mu^*_0}}(Z)
}{p_{\vec{\mu}}(Z)}=\frac{p_{\vec{\mu}}\left(\data\right)}{p_{\repvec{\mu^*_0}}\left(\data\right)}=\Spseudo.\]
Therefore, $\Spseudo$ is also an \E-variable, so we derive that $\Spseudo = \Sripr$ by Theorem \ref{Speudo=Sripr}.
Furthermore, we have that the denominator of $\Smix$ is given by a different distribution than $p_{\repvec{\mu^*_0}}$, so that $\Smix\neq \Sripr=\Spseudo=\Scond$.
The result then follows from \eqref{eq:simplerelation}.

\end{proof}

\subsection{The Families for Which \texorpdfstring{$\Spseudo$}{Spseudo} Is Not an E-variable}
Here, we prove that $\Spseudo$ is not an \E-variable for $\cM$ equal to the family of beta distributions with free $\beta$ and fixed $\alpha$.
It then follows from \eqref{eq:simplerelation} that $\Spseudo \succ \Sripr$.
(\ref{eq:simplerelation}) also gives  $\Sripr \succeq \Smix$ and $ \Sripr \succeq \Scond$. 
The same is true for $\cM$ equal to the family of geometric distributions and the family of Gaussian distributions with free variance and fixed mean, as the proof that $\Spseudo$ is not an \E-variable is entirely analogous to the proof for the beta distributions given below.  
In all of these cases, one easily shows by simulation that in general, $\Sripr \neq \Smix$ and $\Sripr \neq \Scond$, so then $\Sripr \succ \Smix$ and $\Sripr \succ \Scond$ follow.

\begin{proof}\label{beta_Spseudo_Not_Evalue}
First, let $Q_{\alpha,\beta}$ represent a beta distribution in its standard parameterization, so that its density is given by 
$$
q_{\alpha,\beta}(u) = \frac{\Gamma(\alpha+\beta)}{\Gamma(\alpha)\Gamma(\beta)}u^{\alpha-1}(1-u)^{\beta-1}, \qquad \alpha, \beta > 0; u \in [0, 1].
$$
To simplify the proof, we assume $\alpha = 1$ here. Then
$$
q_{1,\beta}(u) = \frac{\Gamma(1+\beta)}{\Gamma(\beta)}(1-u)^{\beta-1} = \frac{1}{1-u}\exp \left(\beta \log(1-u) - \log\frac{1}{\beta} \right) 
$$
where the first equality holds since $\Gamma(1+\beta) = \beta \Gamma(\beta)$. Comparing this to  (\ref{ExponentialFormulaPre}), we see that $\beta$ is the canonical parameter corresponding to the family $\{Q_{1,\beta}: \beta > 0 \}$, and we have  
$$
\lambda(\mu) = \beta, \quad t(u) = \log(1-u), \quad A(\beta) = \log\frac{1}{\beta}.
$$
To prove the statement, according to Proposition \ref{SpseudoAnE-value}, we just need to show, for any $\mu_1,\ldots, \mu_k$ that are not all equal to each other, that, with  $X= t(U) = \log (1-U)$  and  $\mu_0^{*}= \frac{1}{k}\sum\limits_{i=1}^k \mu_i$
defined as in (\ref{eq:mu0star}), we have 
\begin{equation}\label{SimplifiedJudgeE}
\sum\limits_{i=1}^k \text{\sc var}_{P_{\mu_i}}[X] - k \text{\sc var}_{P_{\mu_0^{*}}}[X] > 0.
\end{equation}
Straightforward calculation gives 
\begin{equation}\label{eq:sundaynighta}
\text{\sc var}_{P_{\mu_i}}[X] = \text{\sc var}_{Q_{1,\beta_i}}[X] = \frac{d^2}{d^2 \beta_i} (\log\frac{1}{\beta_i}) = \frac{1}{\beta_i^2}
\text{\ in particular\ }  \text{\sc var}_{P_{\mu_0^{*}}}[X] = \frac{1}{(\beta_0^{*})^2}
\end{equation}
where $\beta_i$ corresponds to $\mu_i$, i.e. $\mathbb{E}_{Q_{1,\beta_i}}\left[(X) \right] = \mu_i$.
We also have: 
\begin{equation}\label{eq:sundaynightb}
\mathbb{E}_{P_{\beta_0^{*}}}\left[(X) \right] = \mu_0^{*} = \frac{1}{k}\sum\limits_{i=1}^k \mu_i = \frac{1}{k}\sum\limits_{i=1}^k \mathbb{E}_{P_{\beta_i}}\left[(X) \right].
\end{equation}
While $\mathbb{E}_{P_{\beta_i}}\left[(X) \right] = \frac{d}{d \beta_i} (\log\frac{1}{\beta_i}) = -\frac{1}{\beta_i}$, therefore $\frac{1}{\beta_0^{*}} = \frac{1}{k}\sum\limits_{i=1}^k \frac{1}{\beta_i}$. We obtain, together with (\ref{eq:sundaynighta}) and (\ref{eq:sundaynightb}), that 
\begin{equation}\label{VarianceDiff}
\sum\limits_{i=1}^k \text{\sc var}_{P_{\mu_i}}\left[(X) \right] - k\text{\sc var}_{P_{\mu_0^{*}}} \left[(X) \right] = \sum\limits_{i=1}^k \frac{1}{(\beta_i)^2} - k \left(\frac{1}{k}\sum\limits_{i=1}^k \frac{1}{\beta_i} \right)^2.
\end{equation}
Jensen's inequality now gives that (\ref{VarianceDiff}) is strictly positive, whenever at least one of the $\mu_i$ is not equal to $\mu^*_0$, which is what we had to show.  
\end{proof}

\section{Graphical Depiction of RIPr-Approximation and Convergence of Li's Algorithm}\label{simulations}
\begin{figure}[ht]
    \centering
    \includegraphics[width=0.45\textwidth]{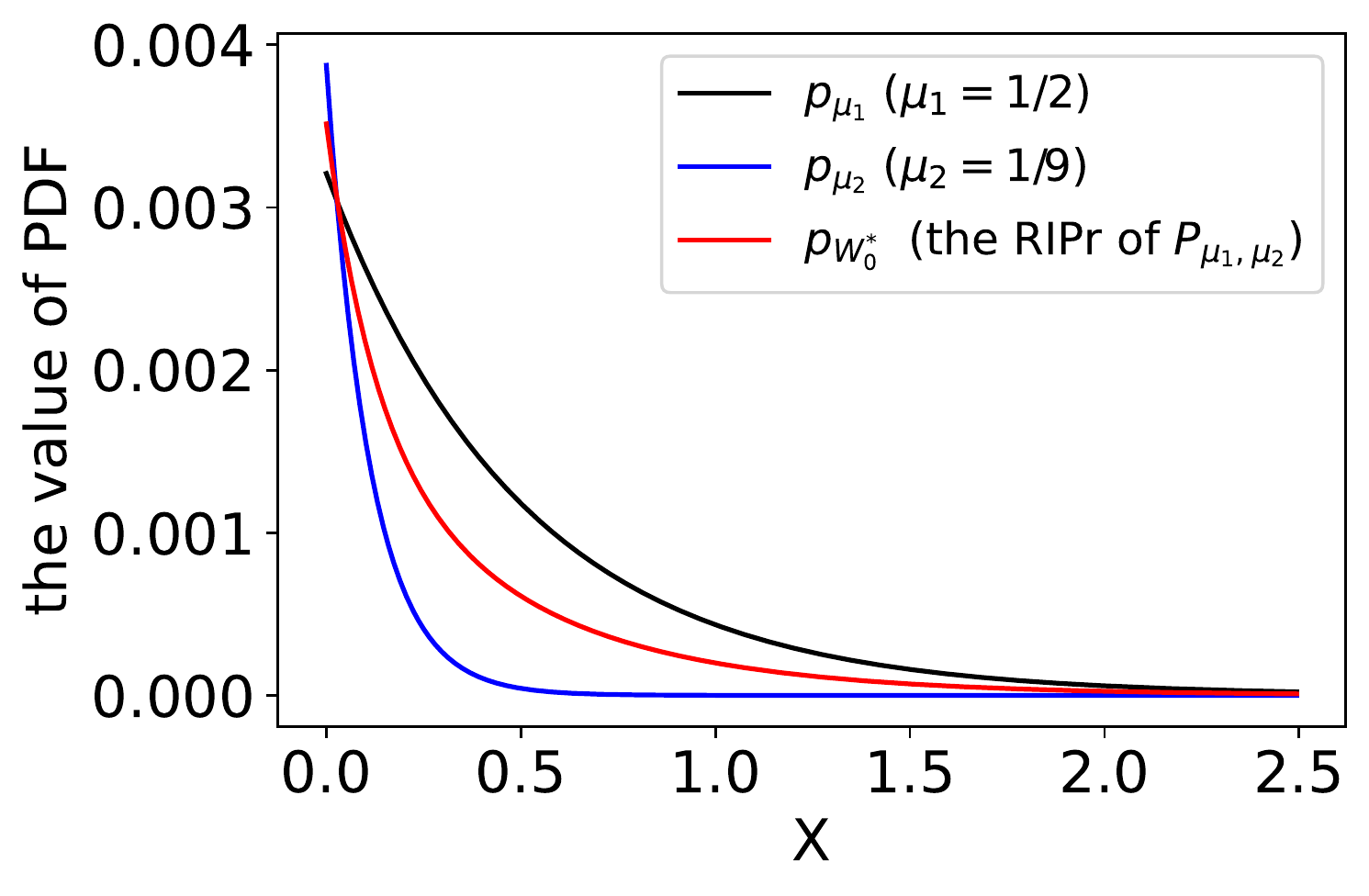}
    \includegraphics[width=0.51\textwidth]{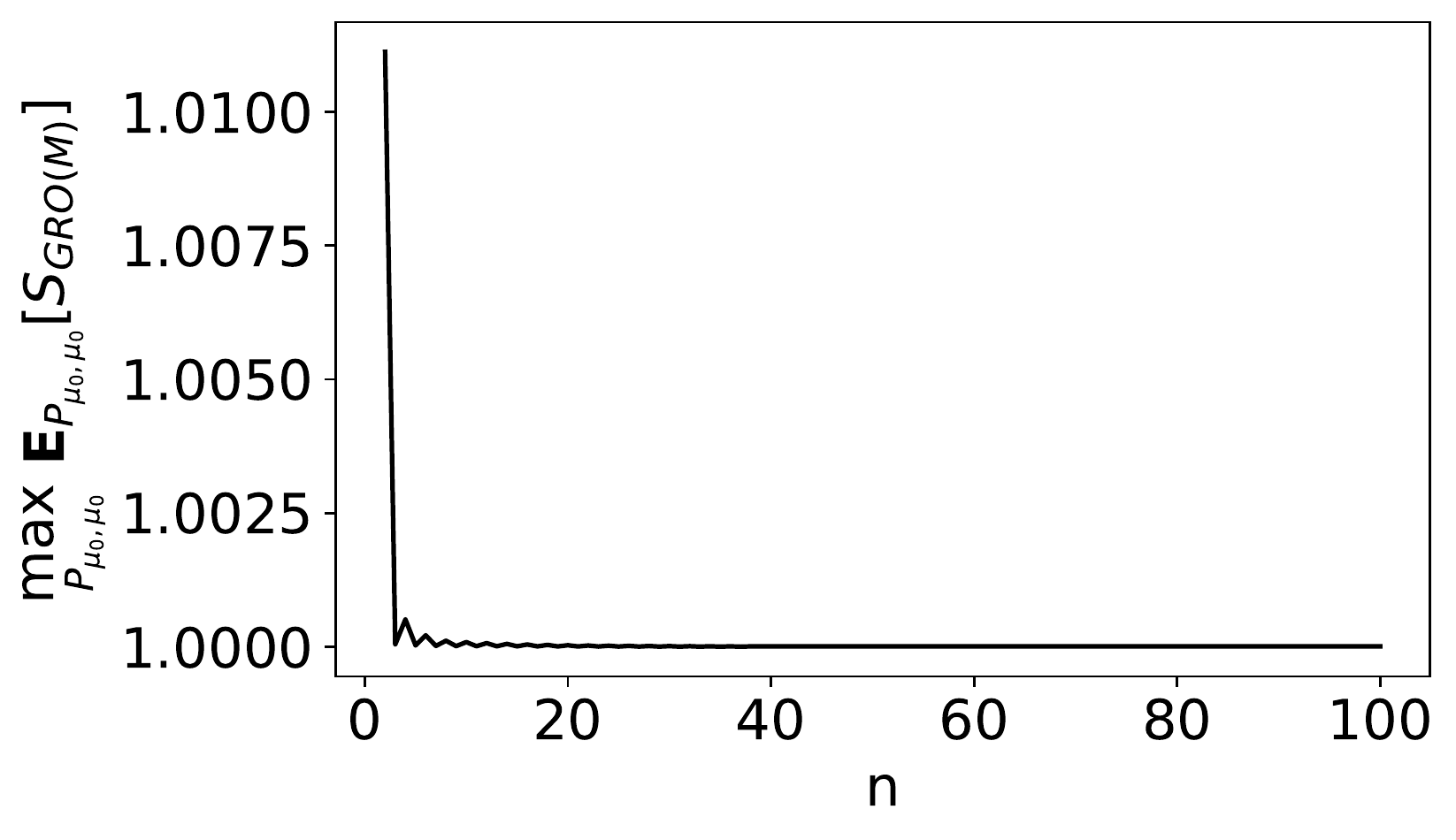}
    \caption{Exponential distribution. On the right, $n$ represents number of iterations with Li's algorithm, starting at iteration 2
    }
    \label{ExpoPDF}
\end{figure}

\begin{figure}[ht]
    \centering
    \includegraphics[width=0.45\textwidth]{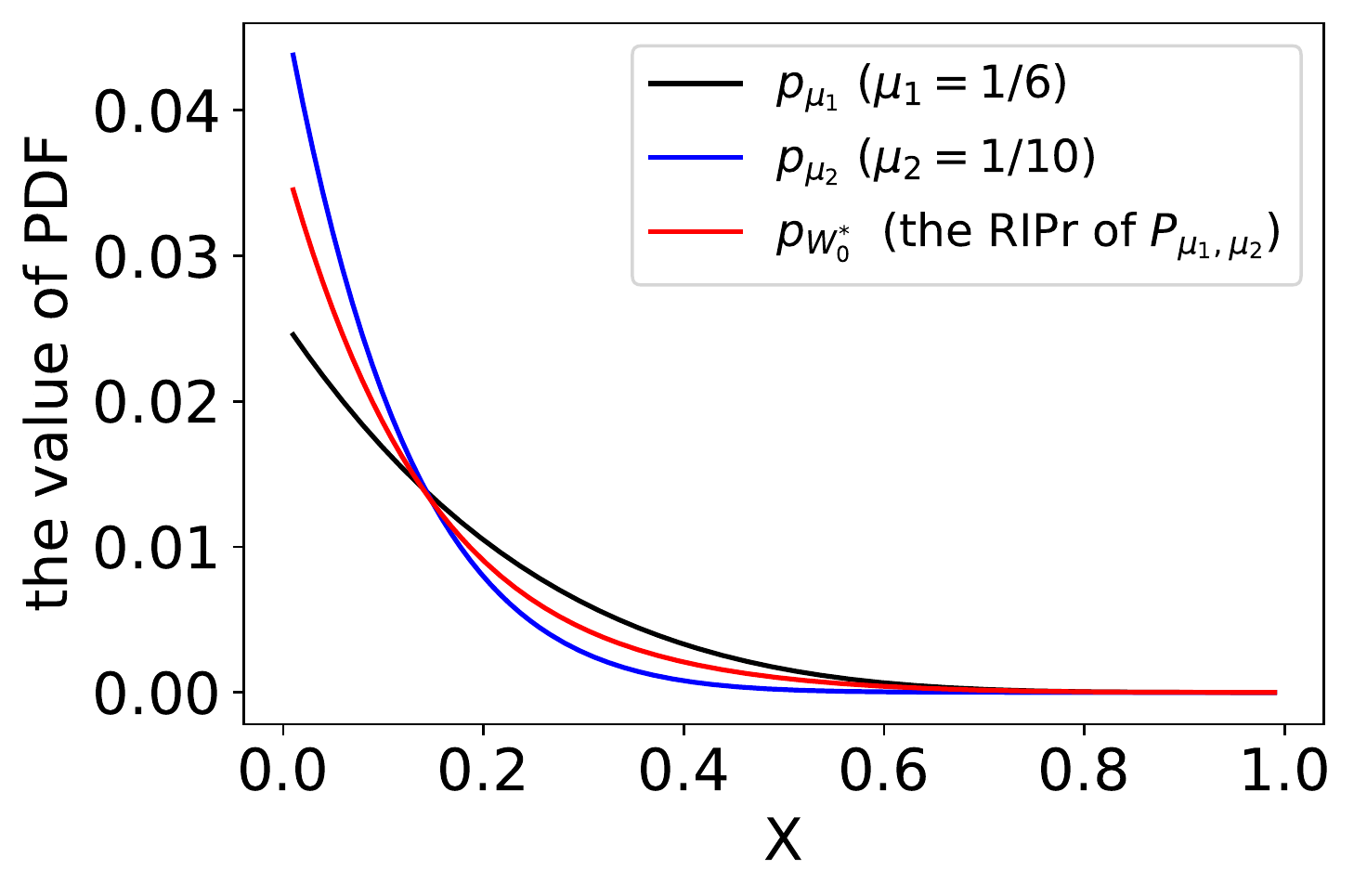}
    \includegraphics[width=0.5\textwidth]{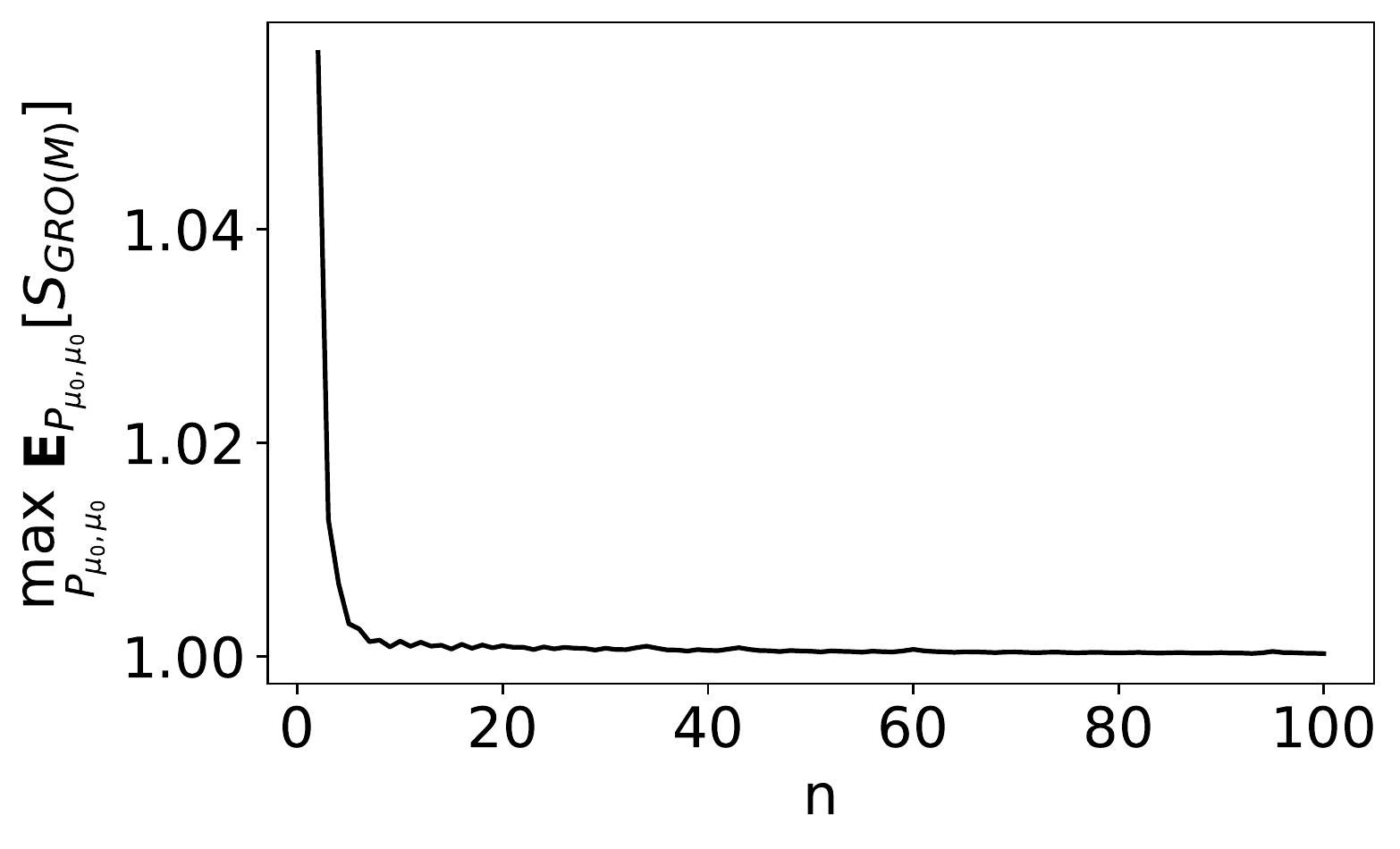}
    \caption{beta with free $\beta$ and fixed $\alpha$. 
    On the right, $n$ represents number of iterations with Li's algorithm, starting at iteration 2
}
    \label{P_mua P_mub P_W0 & li-convergence beta}
\end{figure}

\begin{figure}[ht]
    \centering
    \includegraphics[width=0.45\textwidth]{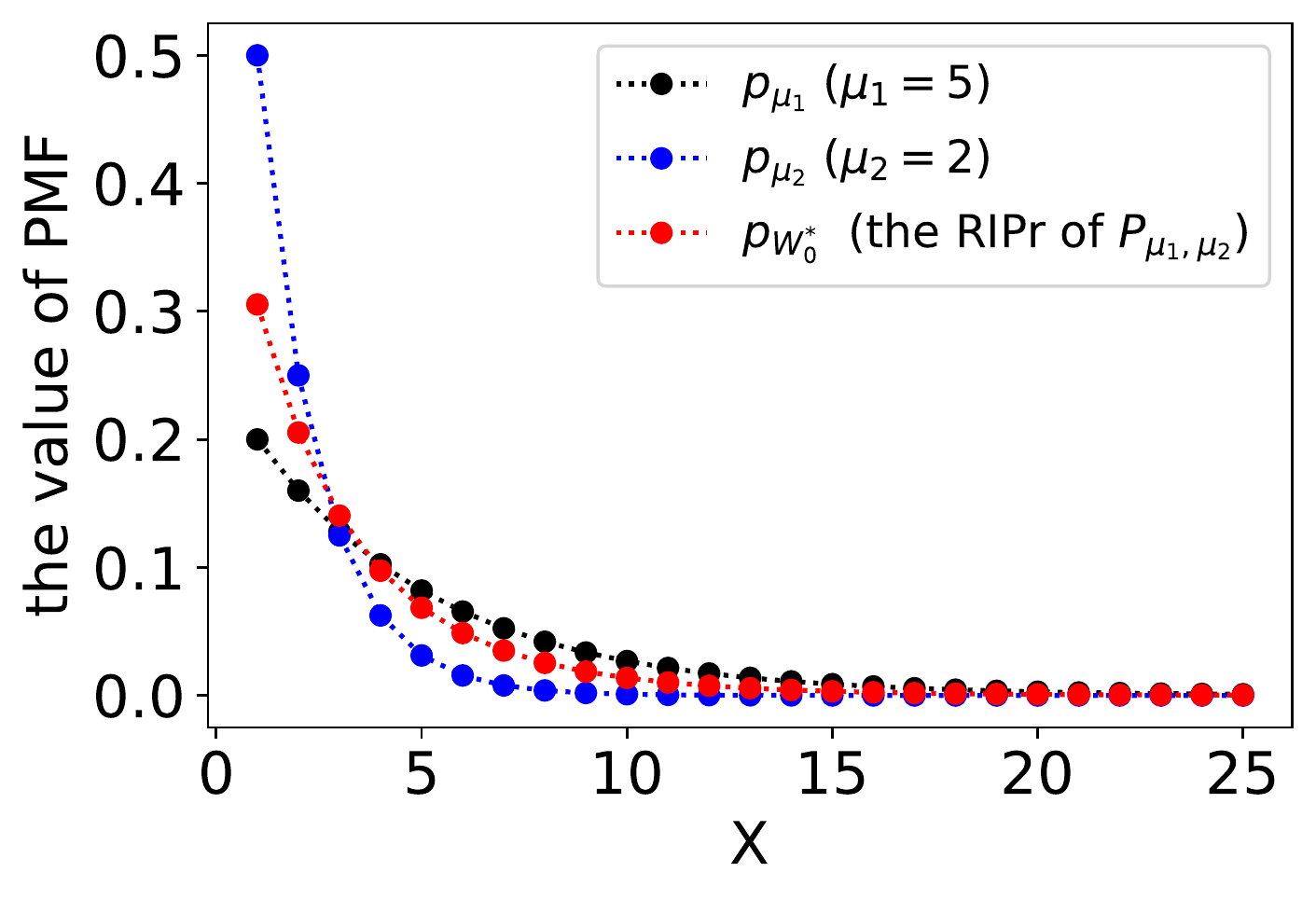}
    \includegraphics[width=0.51\textwidth]{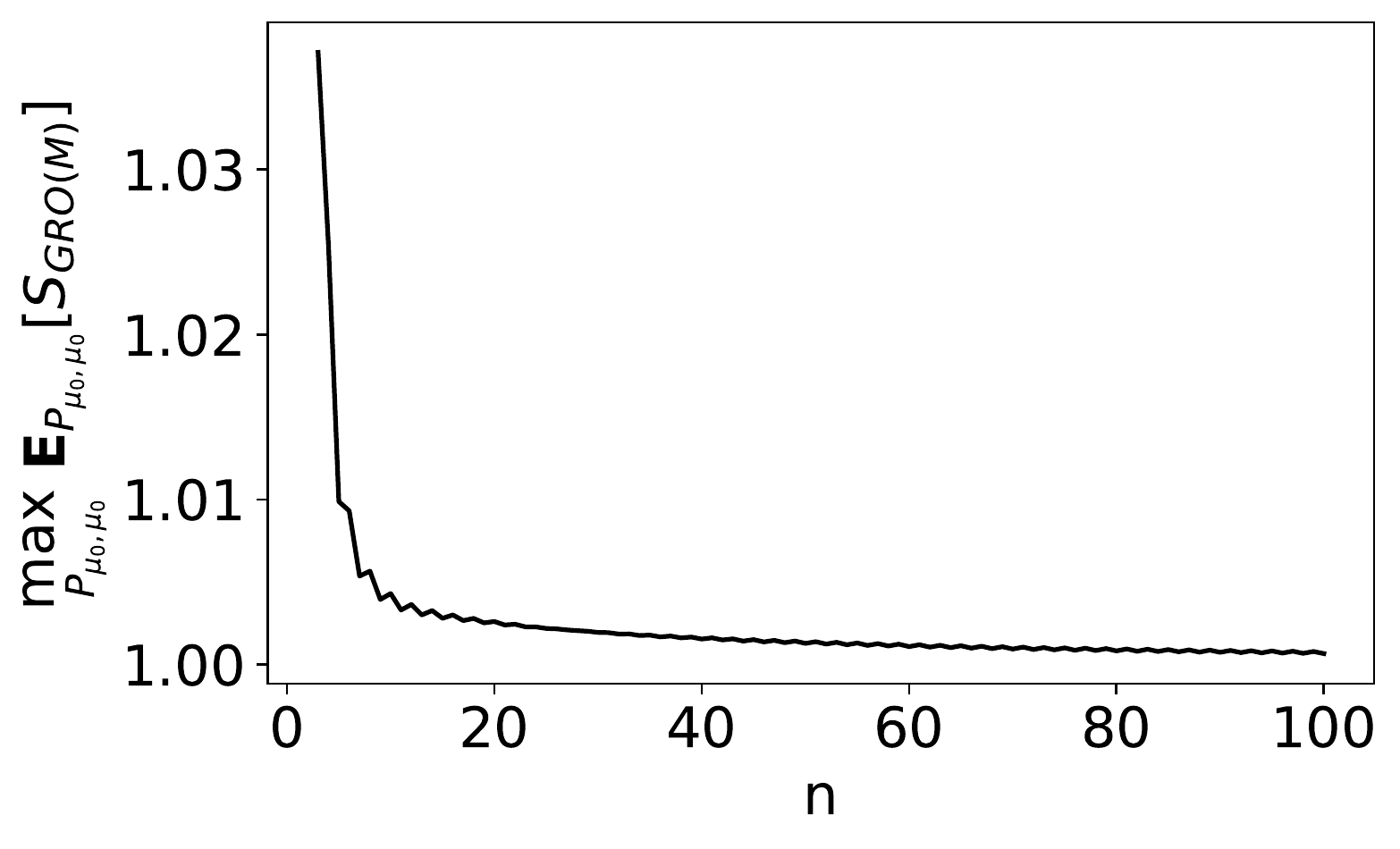}
    \caption{geometric distribution. 
    On the right, $n$ represents number of iterations with Li's algorithm, starting at iteration 3
    }
    \label{geometricPMF}
\end{figure}

\begin{figure}[ht]
    \centering
    \includegraphics[width=0.45\textwidth]{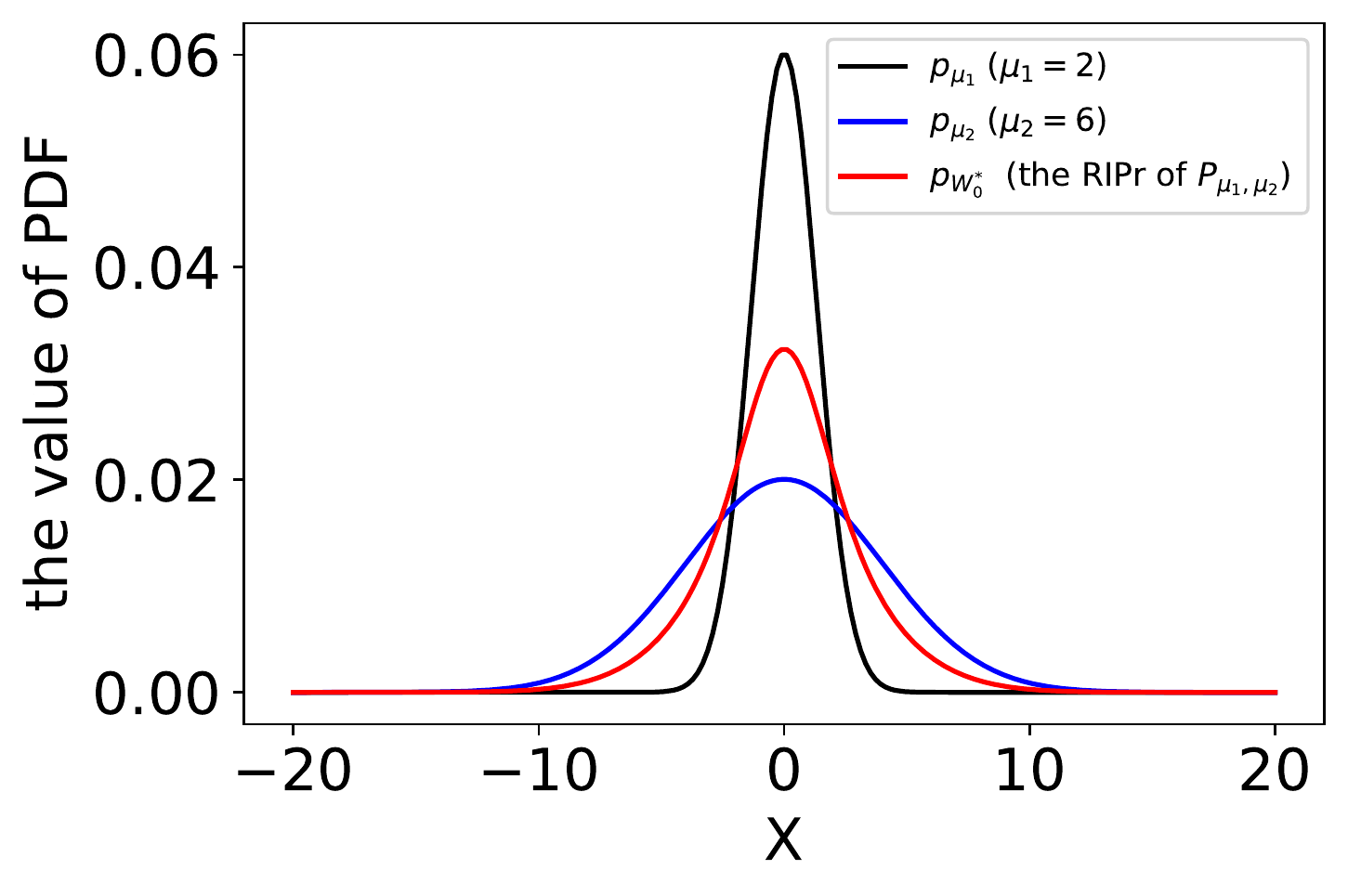}
    \includegraphics[width=0.5\textwidth]{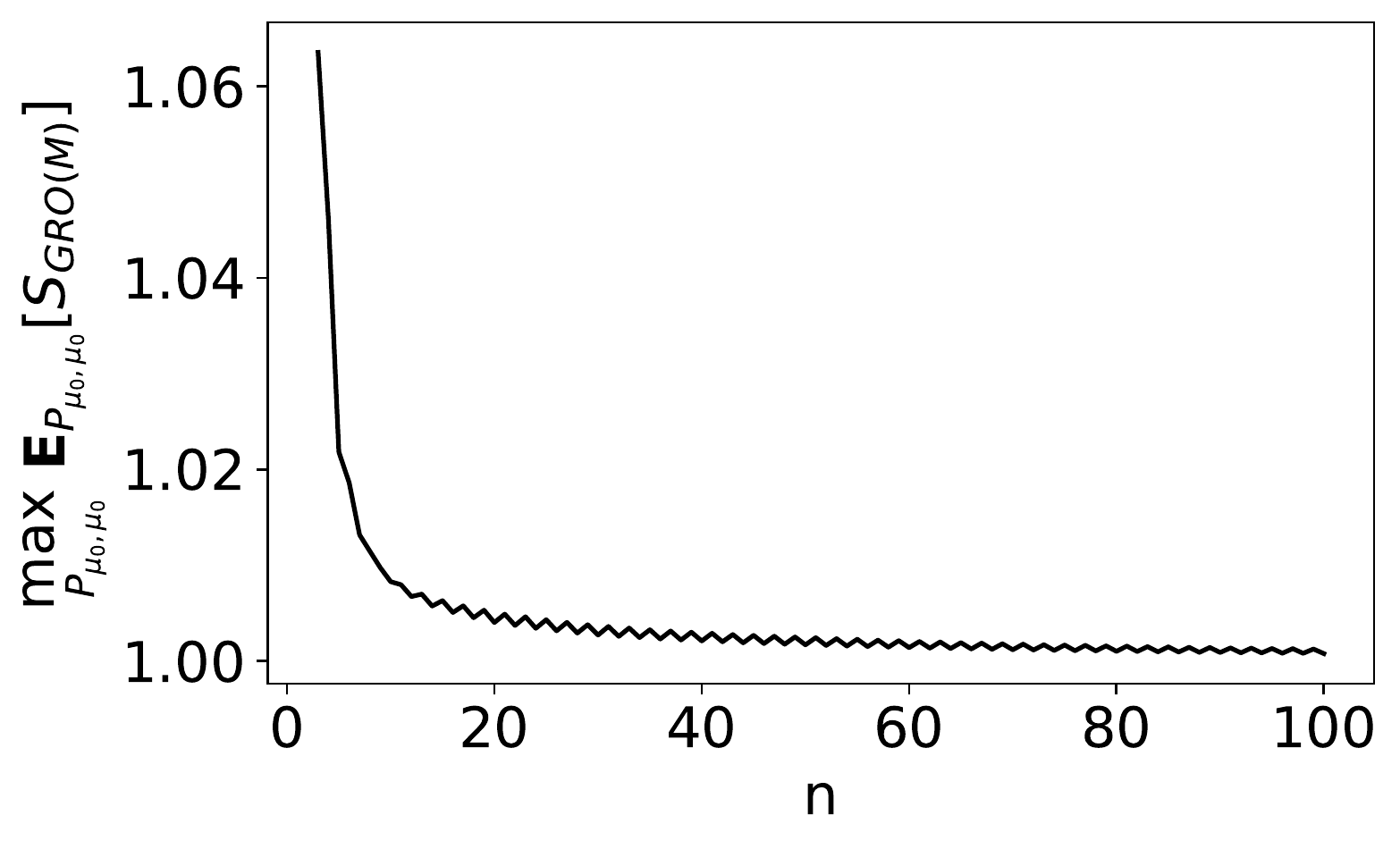}
    \caption{Gaussian with free variance and fixed mean. 
    On the right, $n$ represents number of iterations with Li's algorithm, starting at iteration 3
}
    \label{GaussianPDF}
\end{figure}
We illustrate  RIPr-approximation and convergence of Li's algorithm with four distributions: exponential, beta with free $\beta$ and fixed $\alpha$, geometric and Gaussian with free variance and fixed mean, each with one particular (randomly chosen) setting of the parameters.  The pictures on the left in Figure~\ref{ExpoPDF}--~\ref{GaussianPDF} give the probability density functions (for geometric distributions, discrete probability mass functions) after $n=100$ iterations of Li's algorithm. The pictures on the right illustrate the speed of convergence of  Li's algorithm. 
The pictures on the right do not show the first 
(or the first two, for geometric and Gaussian with free variance) iteration(s), since the worst-case expectation 
$\sup_{\mu_0 \in \mathtt{M}} [\Sripr]$ is invariably incomparably larger in these initial steps. We empirically find that Li's algorithm converges quite  fast for computing the true $\Sripr$. 
In each step of Li's algorithm, we searched for the best mixture weight $\alpha$ in $P_{(m)}$ over a uniformly spaced grid of 100 points in $[0,1]$, and for the novel component $P' = P_{\mu',\mu'}$ by searching for $\mu'$ in a grid of 100 equally spaced points inside the parameter space $\mathtt{M}$ where the left- and right- endpoints of the grid were determined by trial and error. While with this ad-hoc discretization strategy we obviously cannot guarantee any formal approximation results, in practice it invariably worked well:  in all cases, we found that  $\mathop{\text{max}}\limits_{\mu_0 \in \mathtt{M}} \mathbb{E}_{P_{\mu_0, \mu_0}}[\Sripr]$ $\leq$ 1.005 after 15 iterations. For comparison, we show the best approximation that can be obtained by brute-force combining of just two components, for the same parameter values, in Table~\ref{Table_optimal2Components}. 
\begin{table}
    \centering
    \begin{tabular}{ccccc}
    \toprule
    Distributions& $(\mu_1, \mu_2)$& $\alpha$& $(\mu_{01}, \mu_{02})$& $\sup\limits_{\mu_0 \in \mathtt{M}} \mathbb{E}_{X_1, X_2 \sim P_{\mu_0, \mu_0}}[S]$\\
    \midrule
    beta&$(\frac{1}{6}, \frac{1}{10})$& $0.57$& (0.12, 0.16)& 1.00071\\
    geometric&(5, 2)& 0.39& (2.52, 4.21)&  1.00035\\
    Exponential& $(\frac{1}{2}, \frac{1}{9})$& 0.53& (0.13, 0.51)&  1.00083\\
    \makecell{Gaussian with free variance \\ and fixed mean} & (2, 6)& 0.41& (5.82, 3.36)& 1.00035\\
    \bottomrule
    \end{tabular}
    \caption{Analogue of Table~\ref{Table_optimal2Components_samePara} for $\mu_1, \mu_2$ corresponding to the parameters used in Figures~\ref{ExpoPDF}--\ref{GaussianPDF}}
    \label{Table_optimal2Components}
\end{table}
\pagebreak
\section*{Supplementary Material}
In this supplement we verify that all conditions are met for the implicit use of Fubini's theorem and differentiation under the integral sign in the proofs of Theorem 2 and 3, and that all derivatives of interest are bounded.
\subsection*{Theorem 2}
In the paper, notation is as follows:
\begin{align*}
\mu_j &= \mu_0 + \delta \alpha_j \\
\lambda(\mu_j) &= \text{nat. param. $\lambda$ corresponding to mean $\mu = \mu_j$} \\
p_\mu(y) & =  e^{\lambda(\mu) y - A(\lambda(\mu)) } \\
f_y(\delta) & =   \sum_{j=1}^k p_{\mu_i}(y).
\end{align*}
As this will simplify the notation for the derivatives, we write $g_y(\lambda)= e^{\lambda y - A(\lambda)}$, so that 
\begin{equation}\label{eq:crucialrewrite}
f_y(\delta) = \sum_{j=1}^k g_y(\lambda(\mu_j)) \text{\ and\ }
p_{\mu_0}(y) = g_y(\lambda(\mu_0)).
\end{equation}
To stress dependence on $\delta$, we write $\mu_j(\delta)$ instead of $\mu_j$ in the following. 
\paragraph{Step 1}
We first establish the finiteness condition (\ref{eq:finitecheck}).  
We note that 
\begin{align*}
    \log \sum_{j=1}^k g_y(\lambda(\mu_j(\delta))) & \leq \log (\max_j g_y(\lambda(\mu_j(\delta)))k)\\
    &= \max_j \log(g_y(\lambda(\mu_j(\delta)))) + \log k\\
    &\leq \max_j \log(\max\{g_y(\lambda(\mu_j(\delta))),1\}) +\log k\\
    &\leq \sum_j \log(\max\{g_y(\lambda(\mu_j(\delta))),1\}) + \log k \\
    &\leq \sum_j \left| \lambda(\mu_j(\delta))y-\log A(\lambda(\mu_j(\delta))) \right| +\log k.
\end{align*}
and 
\begin{align*}
    \log \sum_{j=1}^k g_y(\lambda(\mu_j(\delta))) &= \log \frac1k \sum_{j=1}^k g_y(\lambda(\mu_j(\delta))) +\log k\\
    &\geq \frac1k \sum_{j=1}^k \log g_y(\lambda(\mu_j(\delta)))+\log k \\
    &= \frac1k \sum_{j=1}^k \lambda(\mu_j(\delta))y- A(\lambda(\mu_j(\delta))) + \log k.
\end{align*}
Putting these together, we see that
\begin{align}\label{eq:crucialfact}
    & |\log f_y(\delta)| \leq \nonumber \\ & \max \left\{ \sum_j \left| \lambda(\mu_j(\delta))y- A(\lambda(\mu_j(\delta))) \right| +\log k, \left| \frac1k \sum_{j=1}^k (\lambda(\mu_j(\delta))y- A(\lambda(\mu_j(\delta)))) + \log k\right| \right\} \nonumber \\
    &\leq \sum_j \left| \lambda(\mu_j(\delta))y- A(\lambda(\mu_j(\delta))) \right| +\log k,
\end{align}
and, more trivially,
\begin{align}\label{eq:crucialfactb}
    |\log g_y(\lambda(\mu_0))| \leq \left| \lambda(\mu_0) y- A(\lambda(\mu_0) \right|.
\end{align}
We know that $\lambda(\mu_j(\delta))$ and $A(\lambda(\mu_j(\delta)))$ are smooth, hence finite functions for $\mu_j(\delta)$ in the interior of the mean-value parameter space ${\mathtt M}$ (see 
\cite[Chapter 9, Theorem 9.1 and Eq. (2)]{BarndorffNielsen78}). 
Since ${\mathtt M}$ is open and for all $j=1..k$, $\mu_j(0)= \mu_0 \in {\mathtt M}$, it follows that $|\log f(y)(\delta) - \log g_y(\lambda(\mu_0))|$ can be written as a smooth, in particular finite function of $|y|$ for all $\delta $ in a compact subset of ${\mathbb R}$ with $0$ in its interior. Since $|y| \leq 1 + y^2$ has finite expectation under all $P_{\mu}$ with $\mu \in \mathtt{M}$, finiteness of (\ref{eq:finitecheck}) follows by (\ref{eq:crucialrewrite}).   

\paragraph{Step 2}
We now proceed to establish that we can differentiate with respect to 
$\delta$ for $\delta$ in a compact subset of ${\mathbb R}$ with $0$ in its interior. The proof will make use of (\ref{eq:crucialfact}) and (\ref{eq:crucialfactb}).
We denote derivatives of functions $f_y$ and $g_y$ as
\[
g^{s}_y(\lambda)  = \frac{\mathrm d^s}{\mathrm d \lambda^s} g_y(\lambda) \quad  \text{and} \quad f^{s}_y(\delta) = \frac{\mathrm{d}^s}{\mathrm{d}\delta^s} f_y(\delta).
\]
We will argue that, for any $s\in \mathbb{N}$, the family $\{\frac{\mathrm d^s}{\mathrm{d}\delta^s}f_y(\delta)\log f_y(\delta)-f_y(\delta)\log g_y(\lambda(\mu_0)):\delta \in \Delta\}$ is uniformly integrable for any compact $\Delta\subset \mathbb{R}$, so that we are allowed to interchange differentiation and integration \cite[see e.g.][Chapter~A16]{williams1991probability}.

Using standard results for exponential families, we have, for $\lambda$ in the interior of the canonical parameter space, 
\begin{align*}
    g^{(1)}_y(\lambda) & =  (y - \mu(\lambda)) g_y(\lambda) \\
    g^{(2)}_y(\lambda) & = - I(\lambda) g_y(\lambda) + (y - \mu(\lambda))^2 g_y(\lambda),
\end{align*}
where $\mu(\lambda)$ denotes the mean-value parameter corresponding to $\lambda$ and $I(\lambda)$ the corresponding Fisher information. 

Continuing this using the fact that $(d^s/d \lambda^s) A(\lambda)$ is continuous for all $s$, gives
\begin{equation}\label{eq:gderiv}
g_y^{(s)}(\lambda) = g_y(\lambda) \cdot h_{y,s}(\lambda) \text{\ with\ } 
h_{y,s}(\lambda) = \sum_{t=1}^s h_{[t,s]}(\lambda) (y- \mu(\lambda))^t
\end{equation}
for some smooth functions $h_{[1,s]}, h_{[2,s]}, \ldots, h_{[s,s]}$ of $\lambda$ (we do not need to know precise definitions of these functions). 
Similarly 
\begin{align*}
    f^{(1)}_y(\delta) = \sum_j g^{(1)}_y(\lambda_{\mu_j(\delta)}) \cdot (\lambda(\mu_j(\delta)))' 
\end{align*}
where $\lambda(\mu_j(\delta))'=\frac{\mathrm d}{\mathrm d\delta} \lambda(\mu_j(\delta))$. We know that $\lambda'(\mu_j(\delta))$ and further derivatives are smooth functions for $\mu_j(\delta)$ in the interior of the mean-value parameter space ${\mathtt M}$ (see 
\cite[Chapter 9, Theorem 9.1 and Eq. (2)]{BarndorffNielsen78}). 
Since this space is open and for all $j=1..k$, $\mu_j(0)= \mu_0 \in {\mathtt M}$, it follows that  $\lambda'(\mu_j(\delta))$ are smooth functions of $\delta$ for $\delta$ in a compact subset of ${\mathbb R}$ with $0$ in its interior.   
Thus, analogously to what we did above with $g^{(s)}$, we get that 
\begin{equation}\label{eq:fderiv}
    f^{(s)}_y(\delta)  = \sum_j \sum_{t=1}^s g^{(t)}_y(\lambda(\mu_j(\delta))) \cdot r_{t,s}(\mu_j) 
\end{equation}
for some smooth functions $r_{t,s}$, the details of which we do not need to know. 
In particular this gives, with 
$$
 b^{(s)}_y:=  \frac{f^{(s)}_y(\delta)}{f_y(\delta)} $$
that 
\begin{align*}
\left|    b^{(s)}_y\right| & 
    \leq 
    \frac{ \sum_j g_y(\lambda(\mu_j(\delta))) \cdot \left( 
    \sum_{t=1}^s |h_{y,t}(\lambda(\mu_j(\delta))) \cdot r_{t,s}(\mu_j(\delta))|  \right) }{\sum_j g_y(\lambda(\mu_j(\delta)))}  \\
    & 
    \leq \sum_j  \sum_{t=1}^s |h_{y,t}(\lambda(\mu_j(\delta))) \cdot r_{t,s}(\mu_j(\delta))|.
\end{align*}
Inspecting the proof in the main text, we informally note that all terms without logarithms in the first four derivatives of $F_0(\delta)$ and $F_1(\delta)$ can be written as
products $f_y(\delta) \cdot b^{(s_1)}_y(\delta) \cdot \ldots \cdot b^{(s_u)}_y(\delta)$ for the $b^{(s)}_y$ we just bounded in terms of polynomials in $|y|$; similarly, the terms involving logarithms can be bounded in terms of such polynomials as well  using (\ref{eq:crucialfact}) and (\ref{eq:crucialfactb}), suggesting that all terms inside all integrals can be such bounded. This is indeed the case: formalizing the reasoning, we see that
\begin{align*}
    &\int \left(\frac{\mathrm d^s}{\mathrm d\delta^s} f_y(\delta)\log f_y(\delta)-f_y(\delta)\log g_y(\lambda(\mu_0))\right)^2 d\rho(y) =\\ & \int \left(f_y^{(s)}(\log f_y(\delta) -\log g_y(\lambda(\mu_0)))+ f_y(\delta) \sum_u c_u\cdot b_y^{(s_2)}(\delta)\cdot \ldots \cdot b_y ^{(s_u)}(\delta) \right)^2 d\rho(y)\\
    &=\int (f_y^{(s)}(\log f_y(\delta) -\log g_y(\lambda(\mu_0))))^2 + \left(f_y(\delta) \sum_u c_u\cdot  b_y^{(s_1)}(\delta)\cdot \ldots \cdot b_y ^{(s_u)}(\delta) \right)^2\\
    &\qquad +f_y(\delta)f_y^{(s)}(\log f_y(\delta)-\log g_y(\lambda(\mu_0)))\sum_u c_u\cdot b_y^{(s_1)}(\delta)\cdot \ldots \cdot b_y ^{(s_u)}(\delta)  d\rho(y).
\end{align*}
By (\ref{eq:crucialfact}) and (\ref{eq:crucialfactb}) and the bound on $|b_y^{(s)}|$ given above, all the terms within the integral can be bounded by polynomials in $y$ (or $|y|$), so the integral is given by linear functions of moments of $\rho$ and $P_{\mu}$. Therefore, using also that  $\rho$ is itself a probability measure and a member of the exponential family under consideration (equal to $P_{\mu}$ with $\lambda(\mu)= 0$), the integral can be uniformly bounded over $\delta$ in a compact subset of the mean-value parameter space.
It follows that the family $\{\frac{\mathrm d^s}{\mathrm{d}\delta^s}f_y(\delta)\log f_y(\delta)-f_y(\delta)\log g_y(\lambda(\mu_0)):\delta \in \Delta\}$ is uniformly integrable~\cite[see e.g.][Chapter~13.3]{williams1991probability}, so integration and differentiation may be interchanged freely~\cite[see e.g.][Chapter A16]{williams1991probability}. It also follows that the quantity on the right-hand side in the theorem statement is bounded.  
\subsection*{Theorem 3}
As in the proof of Theorem 3, let $f(\delta)=\mathbb{E}_{P_{\vec{\mu}}} \left[\log \frac{p_{\vec{\mu}}(X^k) }{p_{\langle {\mu_0} \rangle} (X^k)} - \log  \frac{p_{\vec{\mu}}(X^k\mid Z) }{p_{\langle {\mu_0} \rangle } (X^k\mid Z)} \right]$.

To validate the proof in the main text we merely need to show that $f(\delta)$ is finite, and that we can interchange differentiation and expectation with respect to $\delta$ in a compact interval containing $\delta=0$. Thus, we want to show that, for any $s\in \mathbb{N}$, we have that 
\[\frac{\mathrm d^s}{\mathrm d\delta^s} f(\delta) = \mathbb{E}\left[\frac{\mathrm d^s}{\mathrm d\delta^s} \left(\log \frac{p_{\vec{\mu}}(X^k) }{p_{\langle {\mu_0} \rangle} (X^k)} - \log  \frac{p_{\vec{\mu}}(X^k\mid Z) }{p_{\langle {\mu_0} \rangle } (X^k\mid Z)}\right) \right]. \]
To show this, first note that both $\mathbb{E}_{P_{\vec{\mu}}} \left[\log \frac{p_{\vec{\mu}}(X^k) }{p_{\langle {\mu_0} \rangle} (X^k)} \right]$ and $\mathbb{E}_{P_{\vec{\mu}}}\left[\log  \frac{p_{\vec{\mu}}(X^k\mid Z) }{p_{\langle {\mu_0} \rangle } (X^k\mid Z)} \mid Z\right]$ are KL divergences between members of exponential families (the fact that conditioning on a sum of sufficient statistics results in a new, derived full exponential family is shown by, for example, \cite{Brown86}), which are finite as long as $\delta$ is in a sufficiently small interval containig $0$ in its interior (since then $\mathtt{\mu}$ is in the interior of the mean-value parameter space).  
This already shows that $f(\delta)$ is finite, and it also allows us to rewrite
$$
f(\delta)=\mathbb{E}_{P_{\vec{\mu}}} 
\left[\log \frac{p_{\vec{\mu}}(X^k) }{p_{\langle {\mu_0} \rangle} (X^k)} \right] - \mathbb{E}_{P_{\vec{\mu}}} \left[ \log  \frac{p_{\vec{\mu}}(X^k\mid Z) }{p_{\langle {\mu_0} \rangle } (X^k\mid Z)} \right].$$
Furthermore, 
\cite[Theorem 2.2]{Brown86} in combination with Theorem 9.1. and Chapter 9, Eq.2. of \cite{BarndorffNielsen78} shows that for any full exponential family, for any finite $k> 0$, the $k$-th derivative of the KL divergence with respect to its first argument, given in the mean-value parameterization, exists, is finite, and can be obtained by differentiating under the integral sign, at any $\mu$ in the interior of the mean-value parameter space. 
We are therefore allowed to interchange expectation and differentiation for such terms separately for all $\delta$ in any compact interval containing $0$.
Thus, starting with the previous display, we can write 
\begin{align*}
    & \frac{\mathrm d^s}{\mathrm d\delta^s} f(\delta)  = \frac{\mathrm d^s}{\mathrm d\delta^s} \mathbb{E}_{P_{\vec{\mu}}} \left[\log \frac{p_{\vec{\mu}}(X^k) }{p_{\langle {\mu_0} \rangle} (X^k)} \right] - \frac{\mathrm d^s}{\mathrm d\delta^s}\mathbb{E}_{P_{\vec{\mu}}}\left[\log  \frac{p_{\vec{\mu}}(X^k\mid Z) }{p_{\langle {\mu_0} \rangle } (X^k\mid Z)} \right]\\
    &= \mathbb{E}_{P_{\vec{\mu}}} \left[\frac{\mathrm d^s}{\mathrm d\delta^s}\log \frac{p_{\vec{\mu}}(X^k) }{p_{\langle {\mu_0} \rangle} (X^k)} \right] - \mathbb{E}_{P_{\vec{\mu}}}\left[\frac{\mathrm d^s}{\mathrm d\delta^s}\log  \frac{p_{\vec{\mu}}(X^k\mid Z) }{p_{\langle {\mu_0} \rangle } (X^k\mid Z)} \right]\\
    &= \mathbb{E}_{P_{\vec{\mu}}} \left[\frac{\mathrm d^s}{\mathrm d\delta^s}\log \frac{p_{\vec{\mu}}(X^k) }{p_{\langle {\mu_0} \rangle} (X^k)} \right] - \mathbb{E}_{P_{\vec{\mu}}}\left[\frac{\mathrm d^s}{\mathrm d\delta^s}\log \frac{p_{\vec{\mu}}(X^k) }{p_{\langle {\mu_0} \rangle} (X^k)}  + \log \frac{p_{\vec{\mu};[Z]}(Z) }{p_{\langle \mu_0 \rangle;[Z]} (Z)} \right]= \\
    & \mathbb{E}_{P_{\vec{\mu}}} \left[\frac{\mathrm d^s}{\mathrm d\delta^s}\log \frac{p_{\vec{\mu}}(X^k) }{p_{\langle {\mu_0} \rangle} (X^k)} \right] - \mathbb{E}_{P_{\vec{\mu}}}\left[\frac{\mathrm d^s}{\mathrm d\delta^s}\log \frac{p_{\vec{\mu}}(X^k) }{p_{\langle {\mu_0} \rangle} (X^k)} \right] + \mathbb{E}_{P_{\vec{\mu}}}\left[\frac{\mathrm d^s}{\mathrm d\delta^s}\log \frac{p_{\vec{\mu};[Z]}(Z) }{p_{\langle \mu_0 \rangle;[Z]} (Z)} \right]\\
    &= \mathbb{E}_{P_{\vec{\mu}}}\left[\frac{\mathrm d^s}{\mathrm d\delta^s}\log \frac{p_{\vec{\mu};[Z]}(Z) }{p_{\langle \mu_0 \rangle;[Z]} (Z)} \right],
\end{align*}
where in the last line we use that all involved terms are finite. This is what we had to show. 
\end{document}